\documentclass[longauth]{aa}
\usepackage{graphicx}
\usepackage{hyperref}
\usepackage{natbib}
\usepackage{float}
\usepackage{longtable}
\usepackage{csvsimple}
\usepackage{color}
\usepackage{textcomp}
\usepackage{caption}
\usepackage{subfigure}
\usepackage[retainorgcmds]{IEEEtrantools}
\usepackage[utf8]{inputenc}
\usepackage[varg]{txfonts}
\usepackage{mathabx}
\usepackage{multirow}
\usepackage{booktabs}
\usepackage{placeins}

\makeatletter
\renewcommand*{\@fnsymbol}[1]{\ensuremath{\ifcase#1\or \bigstar\or \bigstar\bigstar\or \ddagger\or
   \mathsection\or \mathparagraph\or \|\or **\or \dagger\dagger
   \or \ddagger\ddagger \else\@ctrerr\fi}}
\makeatother

\usepackage{hyperref}
\hypersetup{
    colorlinks = true,
    linkcolor = {blue}, 
    citecolor = {blue},
    urlcolor = {blue}
}

\begin{document}

    \title{Investigating the visible phase-curve variability of 55 Cnc e\thanks{This article uses data from the CHEOPS programme ID CH\_PR100006.}$^{,}$\thanks{The raw and detrended photometric time-series data are available in electronic form at the CDS via anonymous ftp to cdsarc.cds.unistra.fr (130.79.128.5) or via \url{}}}
    
    \titlerunning{}
    \author{
E. A. Meier Vald\'es\inst{1} $^{\href{https://orcid.org/0000-0002-2160-8782}{\includegraphics[scale=0.5]{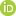}}}$, 
B. M. Morris\inst{2} $^{\href{https://orcid.org/0000-0003-2528-3409}{\includegraphics[scale=0.5]{orcid.jpg}}}$, 
B.-O. Demory\inst{1,3} $^{\href{https://orcid.org/0000-0002-9355-5165}{\includegraphics[scale=0.5]{orcid.jpg}}}$, 
A. Brandeker\inst{4} $^{\href{https://orcid.org/0000-0002-7201-7536}{\includegraphics[scale=0.5]{orcid.jpg}}}$, 
D. Kitzmann\inst{1} $^{\href{https://orcid.org/0000-0003-4269-3311}{\includegraphics[scale=0.5]{orcid.jpg}}}$, 
W. Benz\inst{3,1} $^{\href{https://orcid.org/0000-0001-7896-6479}{\includegraphics[scale=0.5]{orcid.jpg}}}$, 
A. Deline\inst{5}, 
H.-G. Florén\inst{4}, 
S. G. Sousa\inst{6} $^{\href{https://orcid.org/0000-0001-9047-2965}{\includegraphics[scale=0.5]{orcid.jpg}}}$, 
V. Bourrier\inst{5} $^{\href{https://orcid.org/0000-0002-9148-034X}{\includegraphics[scale=0.5]{orcid.jpg}}}$, 
V. Singh\inst{7} $^{\href{https://orcid.org/0000-0002-7485-6309}{\includegraphics[scale=0.5]{orcid.jpg}}}$, 
K. Heng\inst{8,9,10}, 
A. Strugarek\inst{11} $^{\href{https://orcid.org/0000-0002-9630-6463}{\includegraphics[scale=0.5]{orcid.jpg}}}$, 
D. J. Bower\inst{1} $^{\href{https://orcid.org/0000-0002-0673-4860}{\includegraphics[scale=0.5]{orcid.jpg}}}$, 
N. Jäggi\inst{3} $^{\href{https://orcid.org/0000-0002-2740-7965}{\includegraphics[scale=0.5]{orcid.jpg}}}$, 
L. Carone\inst{12} $^{\href{https://orcid.org/0000-0001-9355-3752}{\includegraphics[scale=0.5]{orcid.jpg}}}$, 
M. Lendl\inst{5} $^{\href{https://orcid.org/0000-0001-9699-1459}{\includegraphics[scale=0.5]{orcid.jpg}}}$, 
K. Jones\inst{1}, 
A. V. Oza\inst{13} $^{\href{https://orcid.org/0000-0002-1655-0715}{\includegraphics[scale=0.5]{orcid.jpg}}}$, 
O. D. S. Demangeon\inst{6,14} $^{\href{https://orcid.org/0000-0001-7918-0355}{\includegraphics[scale=0.5]{orcid.jpg}}}$, 
Y. Alibert\inst{3} $^{\href{https://orcid.org/0000-0002-4644-8818}{\includegraphics[scale=0.5]{orcid.jpg}}}$, 
R. Alonso\inst{15,16} $^{\href{https://orcid.org/0000-0001-8462-8126}{\includegraphics[scale=0.5]{orcid.jpg}}}$, 
G. Anglada\inst{17,18} $^{\href{https://orcid.org/0000-0002-3645-5977}{\includegraphics[scale=0.5]{orcid.jpg}}}$, 
J. Asquier\inst{19}, 
T. Bárczy\inst{20} $^{\href{https://orcid.org/0000-0002-7822-4413}{\includegraphics[scale=0.5]{orcid.jpg}}}$, 
D. Barrado Navascues\inst{21} $^{\href{https://orcid.org/0000-0002-5971-9242}{\includegraphics[scale=0.5]{orcid.jpg}}}$, 
S. C. C. Barros\inst{6,14} $^{\href{https://orcid.org/0000-0003-2434-3625}{\includegraphics[scale=0.5]{orcid.jpg}}}$, 
W. Baumjohann\inst{12} $^{\href{https://orcid.org/0000-0001-6271-0110}{\includegraphics[scale=0.5]{orcid.jpg}}}$, 
M. Beck\inst{5} $^{\href{https://orcid.org/0000-0003-3926-0275}{\includegraphics[scale=0.5]{orcid.jpg}}}$, 
T. Beck\inst{3}, 
N. Billot\inst{5} $^{\href{https://orcid.org/0000-0003-3429-3836}{\includegraphics[scale=0.5]{orcid.jpg}}}$, 
X. Bonfils\inst{22} $^{\href{https://orcid.org/0000-0001-9003-8894}{\includegraphics[scale=0.5]{orcid.jpg}}}$, 
L. Borsato\inst{23} $^{\href{https://orcid.org/0000-0003-0066-9268}{\includegraphics[scale=0.5]{orcid.jpg}}}$, 
C. Broeg\inst{3,1} $^{\href{https://orcid.org/0000-0001-5132-2614}{\includegraphics[scale=0.5]{orcid.jpg}}}$, 
J. Cabrera\inst{24}, 
S. Charnoz\inst{25} $^{\href{https://orcid.org/0000-0002-7442-491X}{\includegraphics[scale=0.5]{orcid.jpg}}}$, 
A. Collier Cameron\inst{26} $^{\href{https://orcid.org/0000-0002-8863-7828}{\includegraphics[scale=0.5]{orcid.jpg}}}$, 
Sz. Csizmadia\inst{24} $^{\href{https://orcid.org/0000-0001-6803-9698}{\includegraphics[scale=0.5]{orcid.jpg}}}$, 
P. E. Cubillos\inst{27,12}, 
M. B. Davies\inst{28} $^{\href{https://orcid.org/0000-0001-6080-1190}{\includegraphics[scale=0.5]{orcid.jpg}}}$, 
M. Deleuil\inst{29} $^{\href{https://orcid.org/0000-0001-6036-0225}{\includegraphics[scale=0.5]{orcid.jpg}}}$, 
L. Delrez\inst{30,31} $^{\href{https://orcid.org/0000-0001-6108-4808}{\includegraphics[scale=0.5]{orcid.jpg}}}$, 
D. Ehrenreich\inst{5,32} $^{\href{https://orcid.org/0000-0001-9704-5405}{\includegraphics[scale=0.5]{orcid.jpg}}}$, 
A. Erikson\inst{24}, 
A. Fortier\inst{3,1} $^{\href{https://orcid.org/0000-0001-8450-3374}{\includegraphics[scale=0.5]{orcid.jpg}}}$, 
L. Fossati\inst{12} $^{\href{https://orcid.org/0000-0003-4426-9530}{\includegraphics[scale=0.5]{orcid.jpg}}}$, 
M. Fridlund\inst{33,34} $^{\href{https://orcid.org/0000-0002-0855-8426}{\includegraphics[scale=0.5]{orcid.jpg}}}$, 
D. Gandolfi\inst{35} $^{\href{https://orcid.org/0000-0001-8627-9628}{\includegraphics[scale=0.5]{orcid.jpg}}}$, 
M. Gillon\inst{30} $^{\href{https://orcid.org/0000-0003-1462-7739}{\includegraphics[scale=0.5]{orcid.jpg}}}$, 
M. Güdel\inst{36}, 
M. N. Günther\inst{19} $^{\href{https://orcid.org/0000-0002-3164-9086}{\includegraphics[scale=0.5]{orcid.jpg}}}$, 
S. Hoyer\inst{29} $^{\href{https://orcid.org/0000-0003-3477-2466}{\includegraphics[scale=0.5]{orcid.jpg}}}$, 
K. G. Isaak\inst{19} $^{\href{https://orcid.org/0000-0001-8585-1717}{\includegraphics[scale=0.5]{orcid.jpg}}}$, 
L. L. Kiss\inst{37,38}, 
J. Laskar\inst{39} $^{\href{https://orcid.org/0000-0003-2634-789X}{\includegraphics[scale=0.5]{orcid.jpg}}}$, 
A. Lecavelier des Etangs\inst{40} $^{\href{https://orcid.org/0000-0002-5637-5253}{\includegraphics[scale=0.5]{orcid.jpg}}}$, 
C. Lovis\inst{5} $^{\href{https://orcid.org/0000-0001-7120-5837}{\includegraphics[scale=0.5]{orcid.jpg}}}$, 
D. Magrin\inst{23} $^{\href{https://orcid.org/0000-0003-0312-313X}{\includegraphics[scale=0.5]{orcid.jpg}}}$, 
P. F. L. Maxted\inst{41} $^{\href{https://orcid.org/0000-0003-3794-1317}{\includegraphics[scale=0.5]{orcid.jpg}}}$, 
C. Mordasini\inst{3,1}, 
V. Nascimbeni\inst{23} $^{\href{https://orcid.org/0000-0001-9770-1214}{\includegraphics[scale=0.5]{orcid.jpg}}}$, 
G. Olofsson\inst{4} $^{\href{https://orcid.org/0000-0003-3747-7120}{\includegraphics[scale=0.5]{orcid.jpg}}}$, 
R. Ottensamer\inst{36}, 
I. Pagano\inst{7} $^{\href{https://orcid.org/0000-0001-9573-4928}{\includegraphics[scale=0.5]{orcid.jpg}}}$, 
E. Pallé\inst{15} $^{\href{https://orcid.org/0000-0003-0987-1593}{\includegraphics[scale=0.5]{orcid.jpg}}}$, 
G. Peter\inst{42} $^{\href{https://orcid.org/0000-0001-6101-2513}{\includegraphics[scale=0.5]{orcid.jpg}}}$, 
G. Piotto\inst{23,43} $^{\href{https://orcid.org/0000-0002-9937-6387}{\includegraphics[scale=0.5]{orcid.jpg}}}$, 
D. Pollacco\inst{44}, 
D. Queloz\inst{45,46} $^{\href{https://orcid.org/0000-0002-3012-0316}{\includegraphics[scale=0.5]{orcid.jpg}}}$, 
R. Ragazzoni\inst{23,43} $^{\href{https://orcid.org/0000-0002-7697-5555}{\includegraphics[scale=0.5]{orcid.jpg}}}$, 
N. Rando\inst{19}, 
H. Rauer\inst{24,47,48} $^{\href{https://orcid.org/0000-0002-6510-1828}{\includegraphics[scale=0.5]{orcid.jpg}}}$, 
I. Ribas\inst{17,18} $^{\href{https://orcid.org/0000-0002-6689-0312}{\includegraphics[scale=0.5]{orcid.jpg}}}$, 
N. C. Santos\inst{6,14} $^{\href{https://orcid.org/0000-0003-4422-2919}{\includegraphics[scale=0.5]{orcid.jpg}}}$, 
M. Sarajlic\inst{3}, 
G. Scandariato\inst{7} $^{\href{https://orcid.org/0000-0003-2029-0626}{\includegraphics[scale=0.5]{orcid.jpg}}}$, 
D. Ségransan\inst{5} $^{\href{https://orcid.org/0000-0003-2355-8034}{\includegraphics[scale=0.5]{orcid.jpg}}}$, 
D. Sicilia\inst{7}, 
A. E. Simon\inst{3} $^{\href{https://orcid.org/0000-0001-9773-2600}{\includegraphics[scale=0.5]{orcid.jpg}}}$, 
A. M. S. Smith\inst{24} $^{\href{https://orcid.org/0000-0002-2386-4341}{\includegraphics[scale=0.5]{orcid.jpg}}}$, 
M. Steller\inst{12} $^{\href{https://orcid.org/0000-0003-2459-6155}{\includegraphics[scale=0.5]{orcid.jpg}}}$, 
Gy. M. Szabó\inst{49,50}, 
N. Thomas\inst{3}, 
S. Udry\inst{5} $^{\href{https://orcid.org/0000-0001-7576-6236}{\includegraphics[scale=0.5]{orcid.jpg}}}$, 
B. Ulmer\inst{42}, 
V. Van Grootel\inst{31} $^{\href{https://orcid.org/0000-0003-2144-4316}{\includegraphics[scale=0.5]{orcid.jpg}}}$, 
J. Venturini\inst{5}, 
N. A. Walton\inst{51} $^{\href{https://orcid.org/0000-0003-3983-8778}{\includegraphics[scale=0.5]{orcid.jpg}}}$, 
T. G. Wilson\inst{26} $^{\href{https://orcid.org/0000-0001-8749-1962}{\includegraphics[scale=0.5]{orcid.jpg}}}$, 
D. Wolter\inst{24}
}

    \institute{
\label{inst:1} Center for Space and Habitability, University of Bern, Gesellschaftsstrasse 6, 3012 Bern, Switzerland \and
\label{inst:2} Space Telescope Science Institute, Baltimore, MD 21218, USA \and
\label{inst:3} Physikalisches Institut, University of Bern, Sidlerstrasse 5, 3012 Bern, Switzerland \and
\label{inst:4} Department of Astronomy, Stockholm University, AlbaNova University Center, 10691 Stockholm, Sweden \and
\label{inst:5} Observatoire Astronomique de l'Université de Genève, Chemin Pegasi 51, CH-1290 Versoix, Switzerland \and
\label{inst:6} Instituto de Astrofisica e Ciencias do Espaco, Universidade do Porto, CAUP, Rua das Estrelas, 4150-762 Porto, Portugal \and
\label{inst:7} INAF, Osservatorio Astrofisico di Catania, Via S. Sofia 78, 95123 Catania, Italy \and
\label{inst:8} Ludwig Maximilian University, University Observatory Munich, Scheinerstrasse 1, Munich D-81679, Germany \and
\label{inst:9} University of Warwick, Department of Physics, Astronomy \& Astrophysics Group, Coventry CV4 7AL, United Kingdom \and
\label{inst:10} University of Bern, ARTORG Center for Biomedical Engineering Research, Murtenstrasse 50, CH-3008, Bern, Switzerland \and
\label{inst:11} Université Paris-Saclay, Université Paris Cité, CEA, CNRS, AIM, 91191, Gif-sur-Yvette, France \and
\label{inst:12} Space Research Institute, Austrian Academy of Sciences, Schmiedlstrasse 6, A-8042 Graz, Austria \and
\label{inst:13} Jet Propulsion Laboratory, California Institute of Technology, Pasadena, USA \and
\label{inst:14} Departamento de Fisica e Astronomia, Faculdade de Ciencias, Universidade do Porto, Rua do Campo Alegre, 4169-007 Porto, Portugal \and
\label{inst:15} Instituto de Astrofisica de Canarias, 38200 La Laguna, Tenerife, Spain \and
\label{inst:16} Departamento de Astrofisica, Universidad de La Laguna, 38206 La Laguna, Tenerife, Spain \and
\label{inst:17} Institut de Ciencies de l'Espai (ICE, CSIC), Campus UAB, Can Magrans s/n, 08193 Bellaterra, Spain \and
\label{inst:18} Institut d'Estudis Espacials de Catalunya (IEEC), 08034 Barcelona, Spain \and
\label{inst:19} European Space Agency (ESA), European Space Research and Technology Centre (ESTEC), Keplerlaan 1, 2201 AZ Noordwijk, The Netherlands \and
\label{inst:20} Admatis, 5. Kandó Kálmán Street, 3534 Miskolc, Hungary \and
\label{inst:21} Depto. de Astrofisica, Centro de Astrobiologia (CSIC-INTA), ESAC campus, 28692 Villanueva de la Cañada (Madrid), Spain \and
\label{inst:22} Université Grenoble Alpes, CNRS, IPAG, 38000 Grenoble, France \and
\label{inst:23} INAF, Osservatorio Astronomico di Padova, Vicolo dell'Osservatorio 5, 35122 Padova, Italy \and
\label{inst:24} Institute of Planetary Research, German Aerospace Center (DLR), Rutherfordstrasse 2, 12489 Berlin, Germany \and
\label{inst:25} Université de Paris, Institut de physique du globe de Paris, CNRS, F-75005 Paris, France \and
\label{inst:26} Centre for Exoplanet Science, SUPA School of Physics and Astronomy, University of St Andrews, North Haugh, St Andrews KY16 9SS, UK \and
\label{inst:27} INAF, Osservatorio Astrofisico di Torino, Via Osservatorio, 20, I-10025 Pino Torinese To, Italy \and
\label{inst:28} Centre for Mathematical Sciences, Lund University, Box 118, 221 00 Lund, Sweden \and
\label{inst:29} Aix Marseille Univ, CNRS, CNES, LAM, 38 rue Frédéric Joliot-Curie, 13388 Marseille, France \and
\label{inst:30} Astrobiology Research Unit, Université de Liège, Allée du 6 Août 19C, B-4000 Liège, Belgium \and
\label{inst:31} Space sciences, Technologies and Astrophysics Research (STAR) Institute, Université de Liège, Allée du 6 Août 19C, 4000 Liège, Belgium \and
\label{inst:32} Centre Vie dans l’Univers, Facult\'e des sciences, Universit'e de Gen\`eve, Quai Ernest-Ansermet 30, CH-1211 Gen\`eve 4, Switzerland \and
\label{inst:33} Leiden Observatory, University of Leiden, PO Box 9513, 2300 RA Leiden, The Netherlands \and
\label{inst:34} Department of Space, Earth and Environment, Chalmers University of Technology, Onsala Space Observatory, 439 92 Onsala, Sweden \and
\label{inst:35} Dipartimento di Fisica, Universita degli Studi di Torino, via Pietro Giuria 1, I-10125, Torino, Italy \and
\label{inst:36} Department of Astrophysics, University of Vienna, Türkenschanzstrasse 17, 1180 Vienna, Austria \and
\label{inst:37} Konkoly Observatory, Research Centre for Astronomy and Earth Sciences, 1121 Budapest, Konkoly Thege Miklós út 15-17, Hungary \and
\label{inst:38} ELTE E\"otv\"os Lor\'and University, Institute of Physics, P\'azm\'any P\'eter s\'et\'any 1/A, 1117 Budapest, Hungary \and
\label{inst:39} IMCCE, UMR8028 CNRS, Observatoire de Paris, PSL Univ., Sorbonne Univ., 77 av. Denfert-Rochereau, 75014 Paris, France \and
\label{inst:40} Institut d'astrophysique de Paris, UMR7095 CNRS, Université Pierre \& Marie Curie, 98bis blvd. Arago, 75014 Paris, France \and
\label{inst:41} Astrophysics Group, Keele University, Staffordshire, ST5 5BG, United Kingdom \and
\label{inst:42} Institute of Optical Sensor Systems, German Aerospace Center (DLR), Rutherfordstrasse 2, 12489 Berlin, Germany \and
\label{inst:43} Dipartimento di Fisica e Astronomia "Galileo Galilei", Universita degli Studi di Padova, Vicolo dell'Osservatorio 3, 35122 Padova, Italy \and
\label{inst:44} Department of Physics, University of Warwick, Gibbet Hill Road, Coventry CV4 7AL, United Kingdom \and
\label{inst:45} ETH Zurich, Department of Physics, Wolfgang-Pauli-Strasse 2, CH-8093 Zurich, Switzerland \and
\label{inst:46} Cavendish Laboratory, JJ Thomson Avenue, Cambridge CB3 0HE, UK \and
\label{inst:47} Zentrum für Astronomie und Astrophysik, Technische Universität Berlin, Hardenbergstr. 36, D-10623 Berlin, Germany \and
\label{inst:48} Institut für Geologische Wissenschaften, Freie Universität Berlin, 12249 Berlin, Germany \and
\label{inst:49} ELTE E\"otv\"os Lor\'and University, Gothard Astrophysical Observatory, 9700 Szombathely, Szent Imre h. u. 112, Hungary \and
\label{inst:50} MTA-ELTE Exoplanet Research Group, 9700 Szombathely, Szent Imre h. u. 112, Hungary \and
\label{inst:51} Institute of Astronomy, University of Cambridge, Madingley Road, Cambridge, CB3 0HA, United Kingdom
    }
              
\authorrunning{E.A. Meier Vald\'es et al.}
\date{Received: 2 February 2023 / Accepted: 4 July 2023}

\abstract
{55~Cnc~e is an ultra-short period super-Earth transiting a Sun-like star. Previous observations in the optical range detected a time-variable flux modulation that is phased with the planetary orbital period, whose amplitude is too large to be explained by reflected light and thermal emission alone.}
{The goal of the study is to investigate the origin of the variability and timescale of the phase-curve modulation in 55 Cnc e. To this end, we used the CHaracterising ExOPlanet Satellite (CHEOPS), whose exquisite photometric precision provides an opportunity to characterise minute changes in the phase curve from one orbit to the next.}
{CHEOPS observed 29 individual visits of 55 Cnc e between March 2020 and February 2022. Based on these observations, we investigated the different processes that could be at the origin of the observed modulation. In particular, we built a toy model to assess whether a circumstellar torus of dust driven by radiation pressure and gravity might match the observed flux variability timescale.}
{We find that the phase-curve amplitude and peak offset of 55 Cnc e do vary between visits.  
The sublimation timescales of selected dust species reveal that silicates expected in an Earth-like mantle would not survive long enough to explain the observed phase-curve modulation. We find that silicon carbide, quartz, and graphite are plausible candidates for the circumstellar torus composition because their sublimation timescales are long.}
{The extensive CHEOPS observations confirm that the phase-curve amplitude and offset vary in time. 
We find that dust could provide the grey opacity source required to match the observations.
However, the data at hand do not provide evidence that circumstellar material with a variable grain mass per unit area causes the observed variability.
Future observations with the James Webb Space Telescope (JWST) promise exciting insights into this iconic super-Earth.}

\keywords{Stars: individual: 55 Cnc --
                 Techniques: photometric--
                 Planets and satellites: individual: 55 Cnc e
                 }

\maketitle

\section{Introduction}
\label{section:introduction}

The super-Earth 55~Cnc~e is the only transiting planet of the five planets that are known to be orbiting its star. The star is one of the brightest stars known to host planets ($V=6$). Because of its short orbital period ($P=0.74$ days), 55~Cnc~e is catalogued as an ultra-short period (USP) planet. Among the population of discovered USP planets, 55~Cnc~e is one of the most frequently studied close-in exoplanets. However, the vast number of observations across the entire spectrum, from the ultra-violet (UV) \citep{Bourrier_2018b} to the infrared (IR) \citep{Demory_2011}, did not lead to a conclusive understanding of this object. 

55 Cnc e was discovered via radial velocity (RV) observations at McDonald Observatory with the Hobby-Eberly Telescope (HET) \citep{McArthur_2004} with an RV solution pointing to a 2.808-day period. Later, \citet{Dawson_2010} argued that the reported period was an alias, and they computed a true orbital period of 0.74 days. \citet{Winn_2011} and \citet{Demory_2011} independently discovered the planet to be transiting its host star and confirmed the previously predicted period with the Microvariability and Oscillations of Stars (\textit{MOST}) telescope and the \textit{Spitzer} space telescope, respectively. 

The first photometric observations in the optical revealed that the measured flux at different phases of the planetary orbit could not be explained by thermal emission and reflected light alone \citep{Winn_2011}. An extensive observation campaign between 2011 and 2015 with \textit{MOST} concluded that the phase modulation and phase offset change over time \citep{Dragomir_2012, Sulis_2019}. The occultation was not detected in the \textit{MOST} dataset. \textit{Spitzer} observed 55~Cnc~e multiple times, revealing a significant variability in the occultation depth between 2012 and 2013 \citep{Demory_2015} that was later confirmed by \citet{Tamburo_2018}. However, a recent analysis using the Transiting Exoplanet Survey Satellite (TESS) \citep{Ricker_2015} observations indicated weak evidence of variability in the occultation depth in the optical range across the sectors \citep{Meier_2022}.

The phase-curve of an exoplanet measures the light of a star and planet throughout an orbit. It exposes different sides of the planet to the observer \citep{Knutson_2007, Borucki_2009, Heng_2017}. Between transit and occultation (secondary eclipse), the measured flux varies because a different phase of the planet is observed. Here we focus on tidally locked planets that transit their host star. In the absence of atmospheric dynamics, the flux will reach its lowest point when the planet transits and peak at the secondary eclipse, where the flux corresponds to the star alone. If the phase-curve peak has a phase offset, this could imply atmospheric winds or dynamics. 

Atmospheric dynamics and weather can produce variability in the shape and amplitude of a phase curve, but all other time-varying sources must be ruled out. There is a precedent of observed phase-curve variability of an exoplanet in addition to 55~Cnc~e. \citet{Armstrong_2016} claimed evidence for variability in the atmosphere of the hot Jupiter HAT-P-7 b, but a reassessment by \citet{Lally_2022} concluded that stellar noise might entirely cause the claimed variability. Stars with convective outer envelopes have granules that vary stochastically in time. Supergranulation is a similar dynamical phenomenon that involves horizontal flows that occur on longer timescales, and it displays larger photometric amplitudes \citep{Lally_2022}. In future work, asteroseismic predictions of the variability amplitude at each frequency as a function of the stellar properties may help extrapolate the stellar phenomena of HAT-P-7 (F6V) to 55 Cnc (G8V).

A thermal map of the planet derived with \textit{Spitzer} observations in the IR revealed a hot spot that is offset 41 degrees east of the substellar point \citep{Demory_2016b}. This phase offset can be explained by a narrow region of volcanic activity or by a circulating atmosphere. Moreover, the night-side brightness temperature is about 1400\,K, while the hottest region on the day-side is approximately 1300\,K hotter. The high temperature gradient indicates inefficient heat redistribution from the day-side to the night-side. The temperature contrast and the offset hot spot are consistent with an optically thick atmosphere in which atmospheric recirculation mostly occurs on the day-side, or a planet without atmosphere with magma flows at the surface. The \textit{Spitzer} observations were further analysed by \citet{Angelo_2017}, who concluded that the phase curve favours a substantial atmosphere on 55 Cnc e. A recent reanalysis of the \textit{Spitzer} dataset found a phase offset consistent with zero, with a markedly higher day-side temperature of 3770\,K and a gradient of 2700\,K to the night-side temperature \citep{Mercier_2022}. 

The phase curve of another USP planet was obtained with \textit{Spitzer} and \textit{Kepler}. K2-141 b is a small rocky planet discovered by \textit{Kepler}, orbiting its host star every 6.7 hours. The observations in the IR and optical are consistent with thermal emission and reflected light. The phase offset of this exoplanet is negligible \citep{Zieba_2022}.  

Some effort was expended to identify certain atmospheric species on 55~Cnc~e. So far, there is no evidence of H Ly $\alpha$ absorption \citep{Ehrenreich_2012, Tabernero_2020}, He \citep{Zhang_2021}, H$_{2}$O, TiO \citep{Esteves_2017, Jindal_2020}, CO, CO$_{2}$, HCN, NH$_{3}$, and C$_{2}$H$_{2}$ \citep{Deibert_2021}. \citet{Ridden-Harper_2016} found hints of Ca$_{II}$ H\&K and Na-D, but did not claim detection due to the low significance of the signal and variable Ca$_{II}$. A recent survey \citep{Keles_2022} of a single transit concluded no detection of absorption of O, Si, Al, Na, Mg, K H, P, F, Sr, S, C, Cl, V, Cr, as well as Fe, Ca, Ti, Mn, Ba, Zr, and its singularly ionized forms. This extensive list of no detections strengthens the hypothesis of a heavyweight atmosphere, if any is present at all. 

The process that causes the puzzling observations remains unknown. Temporal variability in UV transit observations suggests star-planet interactions (SPIs) as the possible origin \citep{Bourrier_2018b}. Because of the short orbital distance, the planet and its host star might be magnetically connected \citep{Folsom_2020}. However, more recent work showed that the energy budget seems too low to cause the measured signal \citep{Morris_2021}. Other hypotheses include active volcanism or an inhomogeneous circumstellar dust torus. 

In this paper, we present the results of an extensive campaign with CHEOPS to observe the phase curve of 55~Cnc~e. First, we present the observations, the detrending of the systematics, and the phase-curve model fit to the data in Sect. \ref{section:method}. Then we present the results for the phase-curve amplitude and phase offset in Sect. \ref{section:results}. Based on the results, a discussion follows in Sect. \ref{section:discussion}. We conclude in Sect. \ref{section:conclusions} and highlight future projects that explore this fascinating system. 

\section{Methods}
\label{section:method}

\subsection{CHEOPS}
\label{section:cheops}

The CHaracterising ExOPlanet Satellite (CHEOPS) \citep{Benz_2021} is an on-axis Ritchey-Chr\'etien telescope with a primary mirror with a diameter of 320\,mm. CHEOPS is designed to operate nadir-locked in a Sun-synchronous orbit at an altitude of 700\,km above the Earth's surface. The exposures have a distinctive three-pointed point-spread function (PSF) due to the partial obscuration of the primary mirror by the secondary mirror and its three supports and because the telescope is intentionally defocused \citep{Benz_2021}. The photometer operates in the visible and near-IR range (0.33\,$\mu$m to 1.1\,$\mu$m) using a back-illuminated charge-coupled device (CCD) detector. 

CHEOPS performed 29 photometric visits of 55~Cnc between 23 March 2020 and 26 February 2022, each visit observing at least one orbital period of planet e. The duration of each visit ranged between 25 and 40 hours. Each frame had an exposure time of 44.2 s obtained by stacking 20 individual readout of 2.2 s. The observation log is given in Table \ref{tab:obslog}. The first visit was analysed by \citet{Morris_2021}. Before stacking, small images (called imagettes) of 30 pixels in radius were extracted that contain the PSF of the target star. For each individual frame (referred to as subarray), ten imagettes were downlinked. The observations were reduced with the data reduction pipeline (DRP) \citep{Hoyer_2020}. Complementary to the DRP, \texttt{PIPE} (Brandeker et al., in prep.; see also descriptions in \citeauthor{Brandeker_2022} \citeyear{Brandeker_2022}, \citeauthor{Morris_2021} \citeyear{Morris_2021} and \citeauthor{Szabo_2021} \citeyear{Szabo_2021}) is a photometry-extraction Python package that uses PSF photometry on the 30-pixel imagettes. The results using the subarrays and imagettes are consistent, and thus we chose to use the subarray dataset in this work for computational efficiency. 

\begin{table*}[]
\centering
\caption{CHEOPS observation logs on 55~Cnc.}
\begin{tabular}{cccccccc}
\hline
\hline
Visit & Date Start & Date Stop & File Key & Duration & Integration & Exposures & Efficiency \\
\# & [UTC]& [UTC] & & [hh:mm] & Time [s] & per stack & \%\\ 
\hline
1 & 2020-03-23 13:43 & 2020-03-24 15:59 & {\small CH\_PR100041\_TG000601\_V0200} & 26:15 & 44.0 & 20 ($\times$2.2\,s) & 56 \\
2 & 2020-12-01 14:08 & 2020-12-02 17:25 & {\small CH\_PR100006\_TG000301\_V0200} & 27:16 & 44.2 & 20 ($\times$2.2\,s) & 53 \\
3 & 2020-12-17 05:07 & 2020-12-18 06:48 & {\small CH\_PR100006\_TG000302\_V0200} & 25:40 & 44.2 & 20 ($\times$2.2\,s) & 56 \\
4 & 2020-12-18 07:30 & 2020-12-19 10:31 & {\small CH\_PR100006\_TG000303\_V0200} & 27:01 & 44.2 & 20 ($\times$2.2\,s) & 54 \\
5 & 2020-12-24 18:29 & 2020-12-25 20:08 & {\small CH\_PR100006\_TG000304\_V0200} & 25:40 & 44.2 & 20 ($\times$2.2\,s) & 57 \\
6 & 2020-12-25 20:50 & 2020-12-26 22:31 & {\small CH\_PR100006\_TG000305\_V0200} & 25:41 & 44.2 & 20 ($\times$2.2\,s) & 57 \\
7 & 2020-12-27 19:44 & 2020-12-28 22:19 & {\small CH\_PR100006\_TG000306\_V0200} & 26:35 & 44.2 & 20 ($\times$2.2\,s) & 57 \\
8 & 2021-01-10 06:14 & 2021-01-11 07:51 & {\small CH\_PR100006\_TG000307\_V0200} & 25:37 & 44.2 & 20 ($\times$2.2\,s) & 60 \\
9 & 2021-01-11 08:42 & 2021-01-12 10:14 & {\small CH\_PR100006\_TG000308\_V0200} & 25:31 & 44.2 & 20 ($\times$2.2\,s) & 60 \\
10 & 2021-01-14 09:56 & 2021-01-15 12:12 & {\small CH\_PR100006\_TG000309\_V0200} & 26:16 & 44.2 & 20 ($\times$2.2\,s) & 58 \\
11 & 2021-01-15 12:23 & 2021-01-16 14:39 & {\small CH\_PR100006\_TG000310\_V0200} & 26:16 & 44.2 & 20 ($\times$2.2\,s) & 59 \\
12 & 2021-01-18 18:45 & 2021-01-19 21:01 & {\small CH\_PR100006\_TG000311\_V0200} & 26:16 & 44.2 & 20 ($\times$2.2\,s) & 59 \\
13 & 2021-01-19 21:13 & 2021-01-21 00:16 & {\small CH\_PR100006\_TG000312\_V0200} & 27:04 & 44.2 & 20 ($\times$2.2\,s) & 60 \\
14 & 2021-01-24 18:51 & 2021-01-25 21:16 & {\small CH\_PR100006\_TG000313\_V0200} & 26:25 & 44.2 & 20 ($\times$2.2\,s) & 59 \\
15 & 2021-01-29 23:24 & 2021-01-31 01:40 & {\small CH\_PR100006\_TG000314\_V0200} & 26:16 & 44.2 & 20 ($\times$2.2\,s) & 59 \\
16 & 2021-02-02 18:26 & 2021-02-03 20:42 & {\small CH\_PR100006\_TG000315\_V0200} & 26:16 & 44.2 & 20 ($\times$2.2\,s) & 59 \\
17 & 2021-02-05 11:57 & 2021-02-06 13:35 & {\small CH\_PR100006\_TG000316\_V0200} & 25:38 & 44.2 & 20 ($\times$2.2\,s) & 61 \\
18 & 2021-02-22 09:31 & 2021-02-23 11:09 & {\small CH\_PR100006\_TG000601\_V0200} & 25:37 & 44.2 & 20 ($\times$2.2\,s) & 61 \\
19 & 2021-03-03 01:23 & 2021-03-04 03:39 & {\small CH\_PR100006\_TG000602\_V0200} & 26:16 & 44.2 & 20 ($\times$2.2\,s) & 58 \\
20 & 2021-03-06 12:13 & 2021-03-07 14:29 & {\small CH\_PR100006\_TG000603\_V0200} & 26:16 & 44.2 & 20 ($\times$2.2\,s) & 58 \\
21 & 2021-03-10 07:02 & 2021-03-11 09:39 & {\small CH\_PR100006\_TG000701\_V0200} & 26:37 & 44.2 & 20 ($\times$2.2\,s) & 57 \\
22 & 2021-03-21 17:40 & 2021-03-22 19:18 & {\small CH\_PR100006\_TG000702\_V0200} & 25:38 & 44.2 & 20 ($\times$2.2\,s) & 57 \\
23 & 2021-12-25 10:31 & 2021-12-26 14:23 & {\small CH\_PR100006\_TG000901\_V0200} & 27:52 & 44.2 & 20 ($\times$2.2\,s) & 55 \\
24 & 2022-01-12 08:45 & 2022-01-13 10:17 & {\small CH\_PR100006\_TG000401\_V0200} & 25:32 & 44.2 & 20 ($\times$2.2\,s) & 60 \\
25 & 2022-01-13 22:28 & 2022-01-15 00:11 & {\small CH\_PR100006\_TG000402\_V0200} & 25:43 & 44.2 & 20 ($\times$2.2\,s) & 60 \\
26 & 2022-01-15 00:50 & 2022-01-16 03:56 & {\small CH\_PR100006\_TG000403\_V0200} & 27:06 & 44.2 & 20 ($\times$2.2\,s) & 60 \\
27 & 2022-01-16 04:07 & 2022-01-17 06:23 & {\small CH\_PR100006\_TG000404\_V0200} & 26:16 & 44.2 & 20 ($\times$2.2\,s) & 59 \\
28 & 2022-02-19 16:24 & 2022-02-21 09:00 & {\small CH\_PR100006\_TG001201\_V0200} & 40:35 & 44.2 & 20 ($\times$2.2\,s) & 60 \\
29 & 2022-02-26 15:03 & 2022-02-28 07:30 & {\small CH\_PR100006\_TG001301\_V0200} & 40:26 & 44.2 & 20 ($\times$2.2\,s) & 59 \\
\hline
\end{tabular}
\label{tab:obslog}
\end{table*}

To prepare our data, we first discarded all observations flagged by \texttt{PIPE} (caused by cosmic rays, contamination from a satellite passing through the field of view, or passage above the South Atlantic Anomaly (SAA)) \citep{Brandeker_2022}. We normalised the flux by dividing all measurements by their median value and removed all points above 3$\sigma$ and below 6$\sigma$ from the median of the absolute deviations. The observations that occur before and after Earth occultation often exhibit a significant offset of the star position on the CCD, which is caused by refraction of the light in the upper Earth atmosphere. Therefore, we masked out all measurements with centroids above 3.5$\sigma$ away from the median centroid. 
Additionally, we removed high-background level above 4$\sigma$ from the median. The background level usually increases before or after Earth occultation. We also determined whether light from the nearby star 53 Cnc might affect the flux measurements by leaking into the aperture. While 53 Cnc falls within the full frame of the CCD of 1024x1024 pixels, the 200-pixel aperture is not affected by leaking flux. The DRP report included in every CHEOPS observation file estimates a mean flux due to nearby stars of approximately 0.001\% relative to the flux of the target and thus contributes negligibly to the light curves. The orbital configuration of CHEOPS means that its field of view (FOV) rotates. Any significant flux contamination would therefore be apparent on the CHEOPS orbit timescales of 100 minutes.

\subsection{Detrending basis vectors}
\label{subsection:detrending}

The spacecraft introduces several systematics to photometry. 
Previous works with CHEOPS datasets have corrected these trends via linear regression \citep[e.g.][]{Morris_2021, Delrez_2021, Jones_2022}, Gaussian process \citep[e.g.][]{Lendl_2020, Bonfanti_2021, Leleu_2021, Deline_2022}, or a PSF detrending method \citep{Parviainen_2022, Wilson_2022}. The last method consists of determining vectors associated with changes in the PSF shape by conducting a principal component analysis (PCA) on the subarray images and selecting the vectors that contribute most for use in the light-curve model. For this work, we performed a correction via polynomial regression.

Because of the orbital configuration of CHEOPS, the field of view of the telescope rotates once per orbit. It is necessary to detrend the flux of the star against the roll angle to ensure that the rotating field of view does not introduce correlated noise. We implemented the effect of the roll angle as the following Fourier series \citep{Deline_2022}:
\begin{IEEEeqnarray*}{rll}
\label{eqn:roll angle}
    \textbf{X}_{roll \ angle} = \sum_{i=1}^{N}a_{i} \cos(i\psi_{i}) + b_{i} \sin(i\psi_{i}) \IEEEyesnumber
,\end{IEEEeqnarray*}
where $\psi$ is the roll angle, and $a_{i}$ and $b_{i}$ are the best-fit weights of the roll angle components to the polynomial fit. We limited the series up to fourth order ($N=4$).

A strong increase or decrease in flux at the beginning of a visit has been observed in many datasets, the so-called "ramp effect". This is presumably caused by the change in pointing orientation from one target to the next and is related to the temperature of the spacecraft recorded by the thermistor readout called ThermFront 2. We added a linear term of this thermistor readout as a basis vector to detrend the flux. For some visits, as detailed in the next paragraph, we deemed it necessary to add a quadratic term of ThermFront 2.

\citet{Morris_2021} performed an exhaustive search through different combinations of basis vectors that are correlated with the flux of 55~Cnc and concluded that the combination of the cosine and sine of the roll angle, a unit vector representing the normalised stellar flux and the ThermFront 2 thermistor readout, are the best set for their study. We also performed a statistical comparison based on the Bayesian information criterion (BIC) \citep{Schwarz_1978} for different combinations of basis vectors in a first fit with a linear regression. The reason for using the BIC and not another more robust information criterion is computational efficiency because the fit was only preliminary for the selection of the basis vectors. By definition, the BIC gives a larger penalty per parameter for large datasets and thus favours simpler models \citep{Gelman_2013}. For our current purpose, it suffices. We considered the same set of basis vectors as \citet{Morris_2021} for all visits. In addition to these vectors, we included the background level, quadratic terms of time and thermistor readout, and harmonics of the cosine and sine of the roll angle of Eq. (\ref{eqn:roll angle}) when the BIC favoured the selection ($\Delta$BIC>10). The set of basis vectors used in each visit is listed in Table \ref{tab:basis vectors} in Appendix \ref{appendix:basis vectors}. We constructed an independent design matrix $\textbf{X}$ for each visit with the selected basis vectors. The basis vector coefficients $\beta$ were obtained via polynomial regression and were stored for a later stage. The uncertainties $\sigma_{\beta}$ on the coefficients were obtained from the covariance matrix. 

\subsection{Phase-curve model}
\label{subsection:analysis}

We modelled the flux variation as a sum of terms,
\begin{IEEEeqnarray*}{rll}
\label{eqn:flux}
    &F& = Tr + Occ + f_{P} \IEEEyesnumber
,\end{IEEEeqnarray*}
where $Tr$ is the transit model, $Occ$ is the occultation model, and $f_{P}$ is the phase-curve model. The transit model was based on \citet{Mandel_2002}, implemented in the \texttt{exoplanet} Python package \citep{Foreman-Mackey_2021}. The occultation function was implemented in \texttt{exoplanet} as an eclipse model without limb darkening. To perform a statistical model comparison, we used multiple functional forms for $f_{P}$. First, we modelled the variation in out-of-transit flux with a simple sinusoidal function with a period matching the orbital period of planet e,
\begin{IEEEeqnarray*}{rll}
\label{eqn:sinusoid}
    &f_{P}& = A \cos\left(\frac{2\pi}{P}t-\phi -\pi \right) \IEEEyesnumber
,\end{IEEEeqnarray*}
where $P$ is the planetary orbital period, and $A$ and $\phi$ are the phase-curve amplitude and offset, respectively. We define the orbital phase on $[-\pi , \pi]$ , where $\phi = \pi$ is at the time of mid-transit and $\phi = 0$ is the occultation. A strong sinusoidal modulation of the flux of 55 Cnc phased with the orbital period of planet e was reported by \citet{Winn_2011} and more recently by \citet{Sulis_2019}. Further studies also used a sinusoidal function, among other models \citep{Demory_2016b, Morris_2021}.  

We also considered a phase function based on the assumption that the planet is a Lambertian sphere that scatters isotropically or emits thermally, defined as
\begin{IEEEeqnarray*}{rll}
\label{eqn:lambert}
    &f_{P}& = aE(\sin(|\theta|)+(\pi-|\theta|\cos(|\theta|)) \IEEEyesnumber
,\end{IEEEeqnarray*}
with $a$ the amplitude, $E$ the occultation function, and $\theta$ the orbital phase. This phase function has no phase offset by construction.

\citet{Hu_2015} presented a semi-analytical model\footnote{For details about the derivation, see Appendix A in \citet{Hu_2015}.} for planets in a circular orbit with asymmetric phase variations. This piecewise-Lambertian sphere is suitable for phase-curve observations in the optical range, constructed with significant reflection or emission between two longitudes. This model adds two parameters corresponding to the longitudes $\xi_{1}$ and $\xi_{2}$. Local longitudes ranging between these values have a lower reflectivity than other longitudes on the planet. Our last functional form for the phase curve is a flat line outside of transit and occultation, which implies a constant baseline flux. We tried two variations of a constant continuum, one fitting for the occultation depth, and another assuming an occultation depth fixed to zero. 

Each CHEOPS visit was analysed individually. Based on previous studies \citep{Nelson_2014, Demory_2016b, Bourrier_2018, Tamburo_2018}, we assumed a circular orbit. We further assumed a quadratic limb-darkening law, where the priors on the coefficients specific to the CHEOPS transmission were retrieved from tables containing computed values by \citet{Claret_2021} using the ATLAS \citep{Castelli_2004} and PHOENIX stellar models \citep{Husser_2013}. For 55~Cnc~A, we used a PHOENIX stellar model with an effective temperature of 5200\,K and surface gravity $\log(g)$ = 4.5 \citep{VonBraun_2011}. In our model, we fitted for the time of the mid-transit $t_{0}$, the impact parameter $b$, the quadratic limb-darkening coefficients $u_{0}$ and $u_{1}$, the planet-to-star radius ratio $R_{p}/R_{\star}$, the occultation depth $\delta_{occ}$, the phase-curve parameters, and the best-fit estimators of the basis vectors selected in Sect. \ref{subsection:detrending} and presented in Table \ref{tab:basis vectors}. To deal with the systematics, we implemented a hierarchical model (also known as multilevel model) \citep[]{Yao_2021, McElreath_2016}, where the parameters themselves are described by parameters called hyperparameters. In particular, the mean $\mu_{H}$ and $\sigma_{H}$ are distributions instead of a fixed numerical value. The priors on $\mu_{H}$ use the information of the polynomial regression made previously, while $\sigma_{H}$ is an exponential distribution. The use of hierarchical models allows for an efficient estimate of the uncertainties in the basis vector coefficients, thus avoiding arbitrary scaling of the covariance matrix uncertainties from the polynomial regression. In Table \ref{tab:Parameter priors} we list detailed information about the priors we used. The transit depth was obtained by two different formulations: As the planet-to-star radius ratio squared and for a given stellar limb-darkening law and impact parameter, we implemented the analytic solutions from \citet{Heller_2019}. In our model, the occultation depth parameter was free to explore negative values as well. We set Gaussian priors on the stellar radius and mass based on \citet{VonBraun_2011}. The model was implemented in a Markov chain Monte Carlo (MCMC) using the no u-turn sampler (NUTS) \citep{Hoffmann_2011}, a variant of Hamiltonian Monte Carlo. We sampled the posterior distributions for the parameters with the probabilistic programming package \texttt{PyMC3} \citep{Salvatier_2016}, with 32\,000 draws and 4000 burn-in iterations. After each run, we checked that the chains were well mixed and that the Gelman-Rubin statistic was below 1.01 for all parameters \citep{Gelman_1992}.

\begin{table}[ht]
\centering\setstretch{1.0}
\caption{List of parameters in the MCMC.}
\begin{tabular}{ccc}
\hline
\hline
Parameters & Units & Priors \\
\hline 
$R_{\star}$ & [$R_{\odot}$] & $\mathcal{N}$(0.943, 0.010)\tablefootmark{a} \\
$M_{\star}$ & [$M_{\odot}$] & $\mathcal{N}$(0.905, 0.015)\tablefootmark{a} \\
$u_{0}$ & - & $\mathcal{N}$(0.58128, 0.1)\tablefootmark{b} \\
$u_{1}$ & - & $\mathcal{N}$(0.13712, 0.1)\tablefootmark{b} \\
$t_{0}$ & [BJD-T$_{0}$\tablefootmark{c}] & $\mathcal{N}$(0, 0.01) \\
$P$ & [days] & 0.73654737\tablefootmark{d} \\
$b$ & - & $\mathcal{N}$(0.39, 0.03)\tablefootmark{d} \\
$R_{p}/R_{\star}$ & - & $\mathcal{U}$[0.00187, 0.187] \\
$\delta_{occ}$ & [ppm] & $\mathcal{U}$[-100, 100] \\
$A$ & [ppm] & $\mathcal{U}$[0, 1000] \\
$\phi$ & [rad] & $\mathcal{U}$[$-\pi$, $\pi$]\tablefootmark{*} \\
\textbf{X} & - & $\mathcal{N}$($\mu_{H}$, $\sigma_{H}$) \\
\hline
Hyperparameters & Units & Hyperpriors \\
\hline
$\mu_{H}$ & - & $\mathcal{N}$($\beta$, $\sigma_{\beta}$) \\
$\sigma_{H}$ & - & \textit{Expon}($\lambda$=1)\\
\hline 

            \hline
\end{tabular}
\tablefoot{
 $\mathcal{N}$ and $\mathcal{U}$ represent a normal and uniform distribution, respectively. \textit{Expon} refers to the exponential distribution. T$_{0}$=2458932.56284 BJD is a reference mid-transit time, $\beta$ are the linear regression coefficients of the systematics basis vectors, and $\sigma_{\beta}$ its uncertainties. Reference:
\tablefoottext{a}{\citet{VonBraun_2011}}
\tablefoottext{b}{\citet{Claret_2021}}
\tablefoottext{c}{\citet{Morris_2021}}
\tablefoottext{d}{\citet{Bourrier_2018}}
\tablefoottext{*}{For visits 5, 6, 13, 17, 21, 23, 24, 25, and 28, the range of the uniform distribution is [$0$, 2$\pi$] to avoid bimodal posterior distributions in the phase offset angle.}
}
\label{tab:Parameter priors}
\vspace{-4mm}
\end{table}

\section{Results}
\label{section:results}

\begin{figure*}[!h]
\centering
\subfigure{\includegraphics[width=0.85\textwidth]{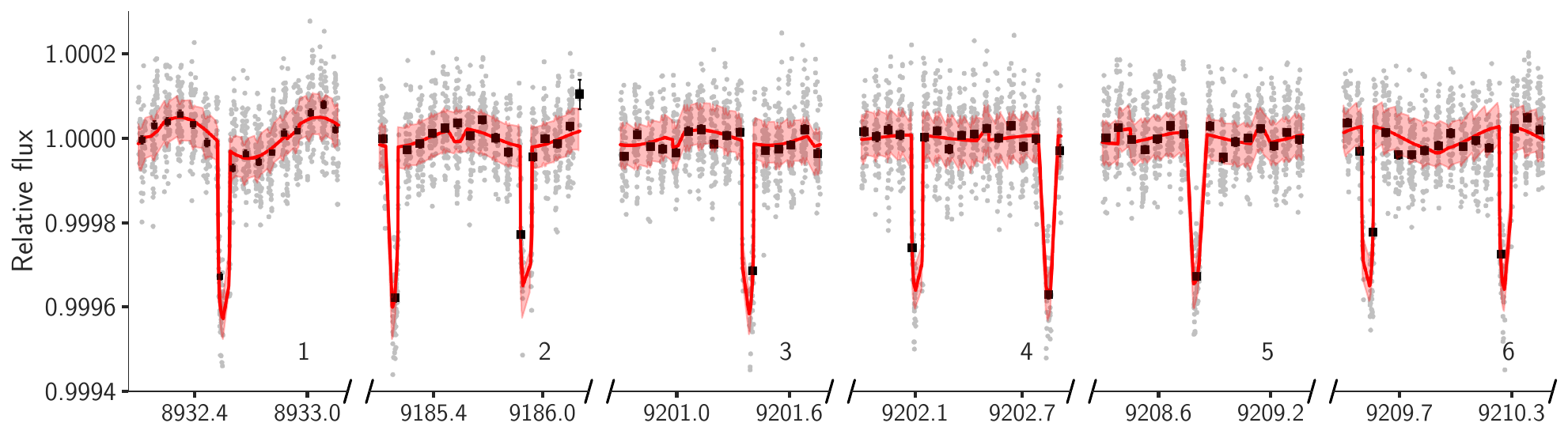}}

\vspace{-5mm}
\subfigure{\includegraphics[width=0.85\textwidth]{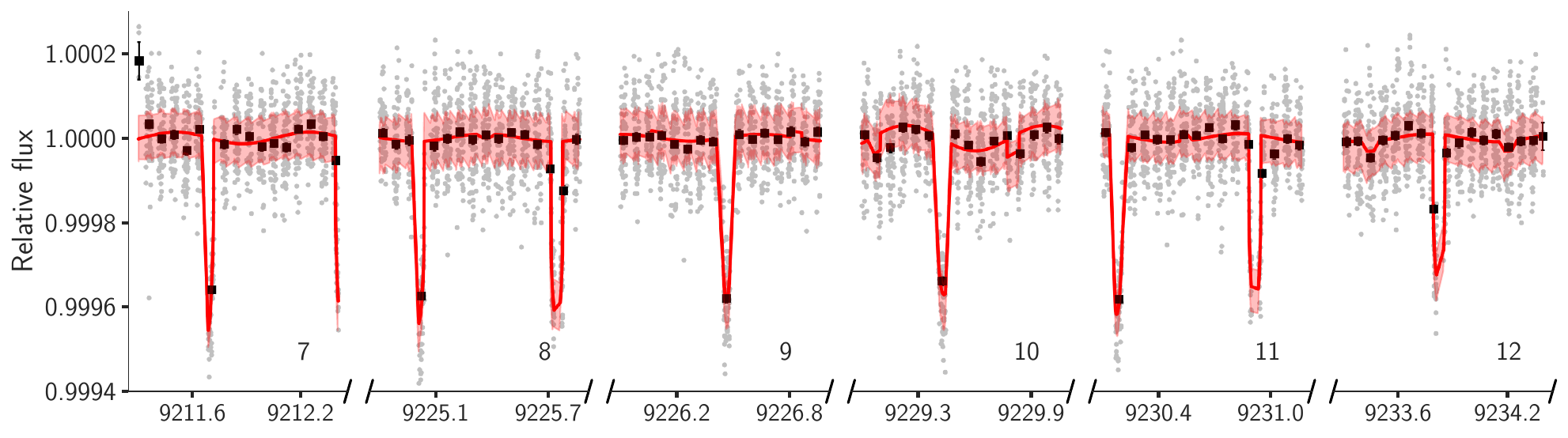}}

\vspace{-5mm}
\subfigure{\includegraphics[width=0.85\textwidth]{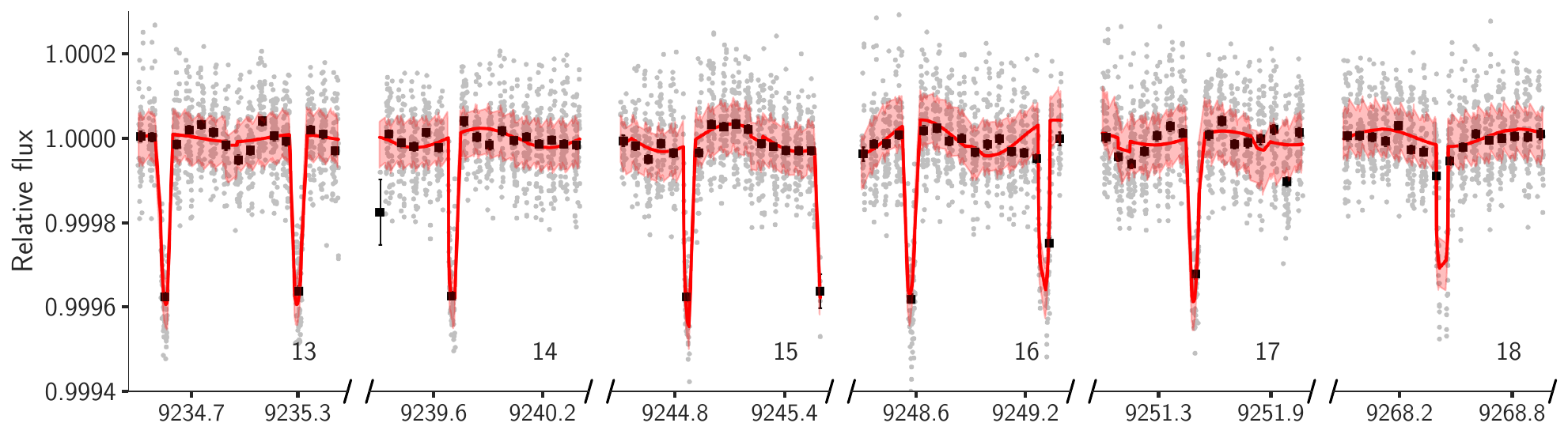}}

\vspace{-5mm}
\subfigure{\includegraphics[width=0.85\textwidth]{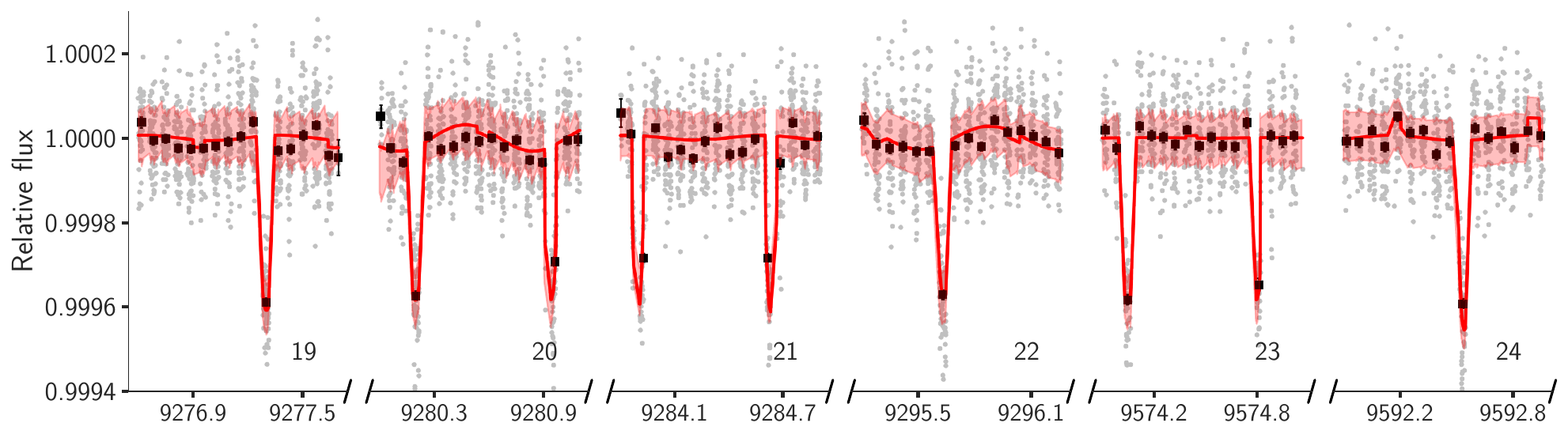}}

\vspace{-5mm}
\subfigure{\includegraphics[width=0.85\textwidth]{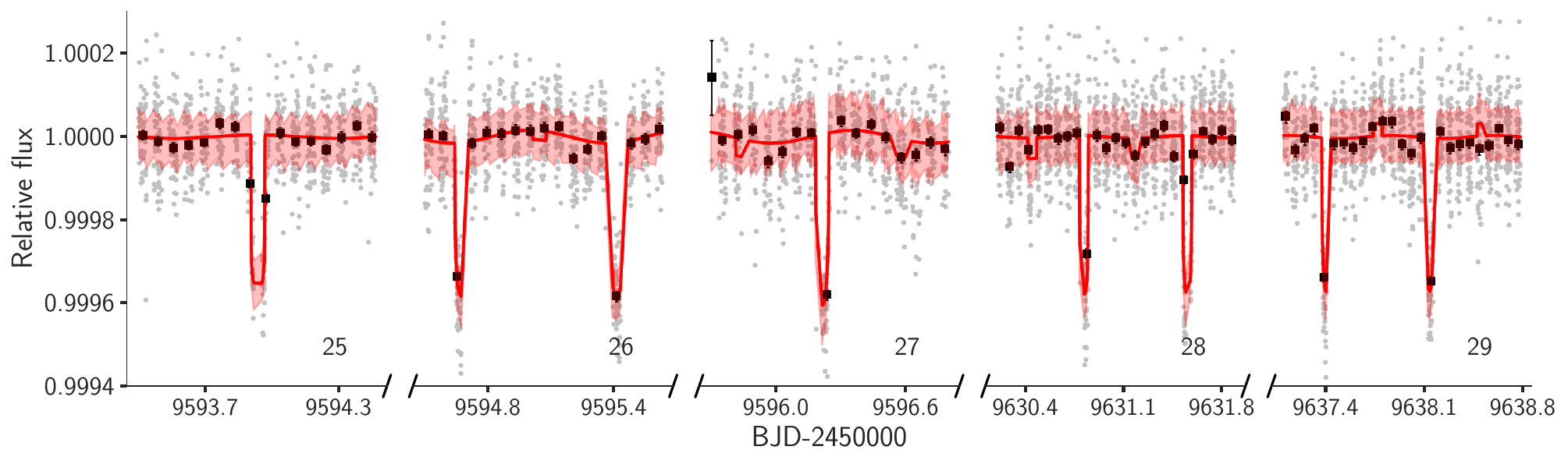}}

\caption{Detrended photometry of 55~Cnc~e for every visit. The detrended flux is plotted in grey, the best-fit phase-curve model in red, and the 1$\sigma$ uncertainty from the posterior distribution in shaded light red. 
The black squares represent the binned data per CHEOPS orbit. The integer close to the bottom right corner of each panel indicates the number of the CHEOPS visit (see Table \ref{tab:obslog}).}
\label{fig:gallery}
\end{figure*}

We present the results assuming a sinusoidal shape for the phase curve based on the fact that this model is preferred in most visits (see Sect. \ref{subsection:model comparison} for details). The detrended flux and samples of the fitted phase-curve model for every CHEOPS visit are shown in a gallery in Fig. \ref{fig:gallery}, while Table \ref{tab:phase curve summary} presents the best-fit values of the phase-curve parameters with $1\sigma$ uncertainty and the residual root mean square (RMS) of each visit. The overlapping phase-curve model and confidence interval in Fig. \ref{fig:gallery} exhibit sharp curves in the transit and spikes due to stitching between the gaps. Appendix \ref{appendix:residuals} presents the residual flux.  

\subsection{Transit and occultation depth}
\label{subsection:transit and occultation depth}

\begin{figure}[ht]
    \centering
    \resizebox{\hsize}{!}{\includegraphics{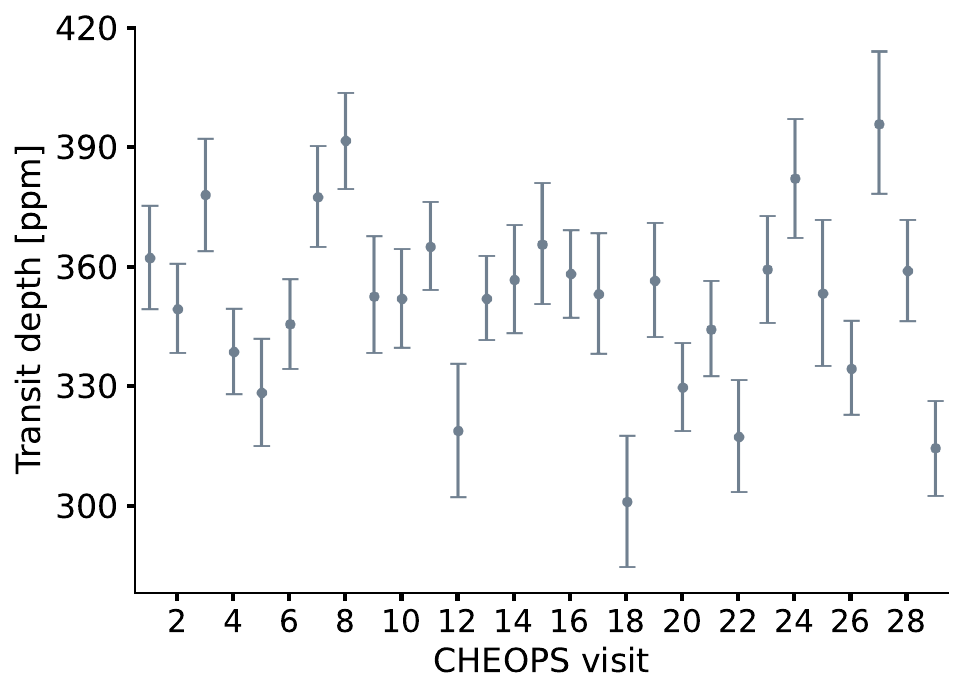}}
    \caption{Transit depth in ppm for each CHEOPS visit. The error bars represent the 1$\sigma$ uncertainty.}
    \label{fig:transit}
\end{figure}

The transit depth across visits, shown in Fig. \ref{fig:transit}, agree within 3$\sigma$ with a mean transit depth of 348 ppm, except for visits 8, 18, and 29. Visual inspection of the residuals of these visits reveals trends that remain in the data, especially in visit 18, which exhibits a strong trend out of transit. In general, these visits are noisier than others, as indicated by the residual RMS. The moderate evidence of variability in the transit depth agrees with \citet{Demory_2015}, \citet{Bourrier_2018}, \citet{Tamburo_2018}, \citet{Sulis_2019}, and \citet{Meier_2022}.

\begin{figure}[ht]
    \centering
    \resizebox{\hsize}{!}{\includegraphics{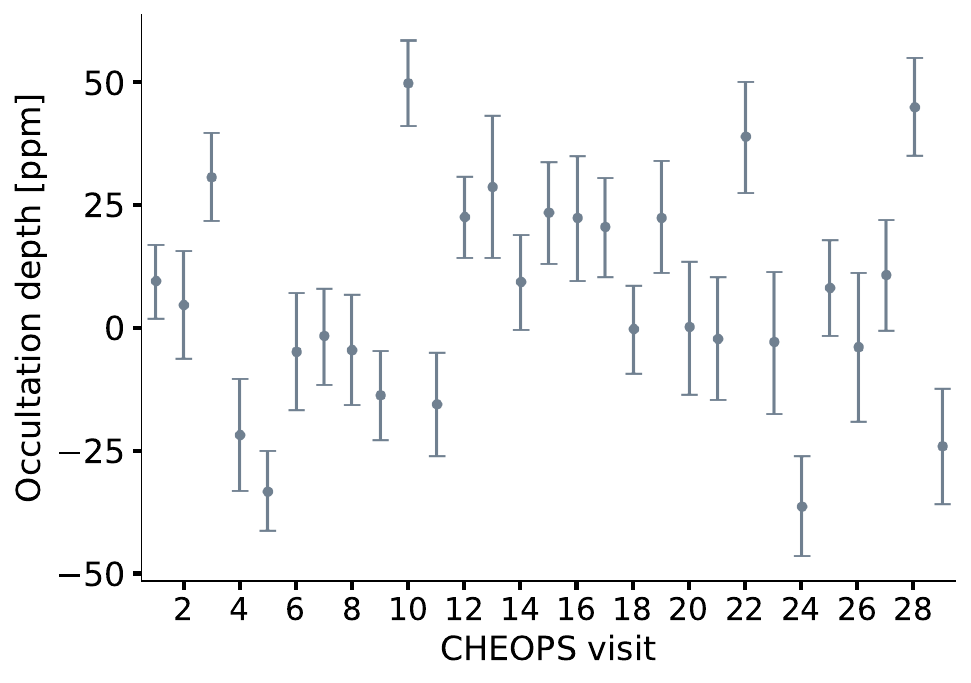}}
    \caption{Occultation depth for each CHEOPS visit. The error bars represent the 1$\sigma$ uncertainty.}
    \label{fig:occultation}
\end{figure}

The occultation depth (Fig. \ref{fig:occultation}) varies by more than 3$\sigma$ in some visits, notably visits 5, 10, 24, and 28. The best-fit occultation depth is negative in many cases, which lacks physical meaning. There seems to be a temporal correlation in the trend between the transit depth and occultation depth, which is especially clear in the first six visits. It is possible that fitting each visit independently for the orbital parameters, such as the impact parameter or mid-transit time, causes this trend. Our results are consistent with those in \citet{Demory_2022}, who used the same CHEOPS dataset. In Appendix \ref{appendix:phase curve parameters} we present plots of the relation between relevant phase-curve parameters.    

\begin{table}[h]
\centering\setstretch{1.5}
\caption{Best-fit phase-curve parameters and residual RMS.}
\begin{tabular}{cccc}
\hline
\hline
Visit & Amplitude [ppm] & Phase offset [$^{\circ}$] & RMS [ppm] \\
\hline 
1 & $50.93_{-3.36}^{+3.41}$ & $59_{-4.6}^{+4.8}$ & $70$ \\
2 & $25.04_{-4.73}^{+4.66}$ & $11_{-11.2}^{+10.3}$ & $71$ \\
3 & $19.93_{-4.21}^{+4.27}$ & $59_{-13.0}^{+12.5}$ & $71$ \\
4 & $7.67_{-4.62}^{+4.79}$ & $119_{-46.5}^{+28.9}$ & $76$ \\
5 & $12.52_{-4.59}^{+4.73}$ & $154_{-20.2}^{+15.9}$ & $71$ \\
6 & $20.06_{-4.40}^{+4.36}$ & $185_{-9.9}^{+9.5}$ & $74$ \\
7 & $11.55_{-4.31}^{+4.11}$ & $82_{-21.3}^{+22.9}$ & $76$ \\
8 & $4.46_{-3.11}^{+5.05}$ & $-12_{-70.9}^{+73.2}$ & $83$ \\
9 & $6.48_{-3.79}^{+3.87}$ & $-94_{-44.8}^{+50.5}$ & $77$ \\
10 & $19.52_{-5.01}^{+4.96}$ & $92_{-13.8}^{+14.8}$ & $78$ \\
11 & $8.06_{-3.93}^{+3.89}$ & $119_{-43.4}^{+32.5}$ & $75$ \\
12 & $10.90_{-4.79}^{+4.77}$ & $134_{-28.1}^{+20.1}$ & $80$ \\
13 & $9.39_{-5.37}^{+5.56}$ & $194_{-23.7}^{+24.6}$ & $80$ \\
14 & $16.06_{-3.98}^{+4.06}$ & $-95_{-15.2}^{+16.7}$ & $78$ \\
15 & $32.92_{-5.17}^{+5.07}$ & $-66_{-8.1}^{+8.3}$ & $80$ \\
16 & $21.87_{-5.29}^{+5.82}$ & $-124_{-15.6}^{+19.5}$ & $84$ \\
17 & $27.25_{-5.32}^{+5.35}$ & $-154_{-8.8}^{+9.7}$ & $89$ \\
18 & $13.75_{-6.78}^{+6.31}$ & $4_{-17.3}^{+17.9}$ & $82$ \\
19 & $9.95_{-6.15}^{+6.14}$ & $124_{-33.9}^{+35.3}$ & $85$ \\
20 & $16.36_{-5.36}^{+5.37}$ & $-47_{-17.7}^{+15.1}$ & $88$ \\
21 & $7.39_{-4.85}^{+5.33}$ & $185_{-35.9}^{+32.9}$ & $81$ \\
22 & $32.41_{-6.01}^{+6.04}$ & $-35_{-11.3}^{+11.4}$ & $88$ \\
23 & $5.80_{-4.07}^{+5.97}$ & $206_{-64.3}^{+47.1}$ & $83$ \\
24 & $5.22_{-3.62}^{+5.35}$ & $256_{-146.6}^{+67.5}$ & $83$ \\
25 & $6.53_{-4.36}^{+5.19}$ & $120_{-55.5}^{+50.5}$ & $83$ \\
26 & $17.38_{-5.34}^{+5.58}$ & $-27_{-17.9}^{+15.0}$ & $82$ \\
27 & $28.65_{-7.50}^{+7.46}$ & $-145_{-11.0}^{+12.2}$ & $89$ \\
28 & $2.91_{-2.05}^{+3.25}$ & $158_{-91.1}^{+109.5}$ & $86$ \\
29 & $5.29_{-3.62}^{+4.81}$ & $15_{-48.3}^{+45.0}$ & $87$ \\
            \hline
        \end{tabular}
        \label{tab:phase curve summary}
        \vspace{-4mm}
\end{table}

\subsection{Amplitude}
\label{subsection:amplitude}
\begin{figure}[h]
    \centering
    \resizebox{\hsize}{!}{\includegraphics{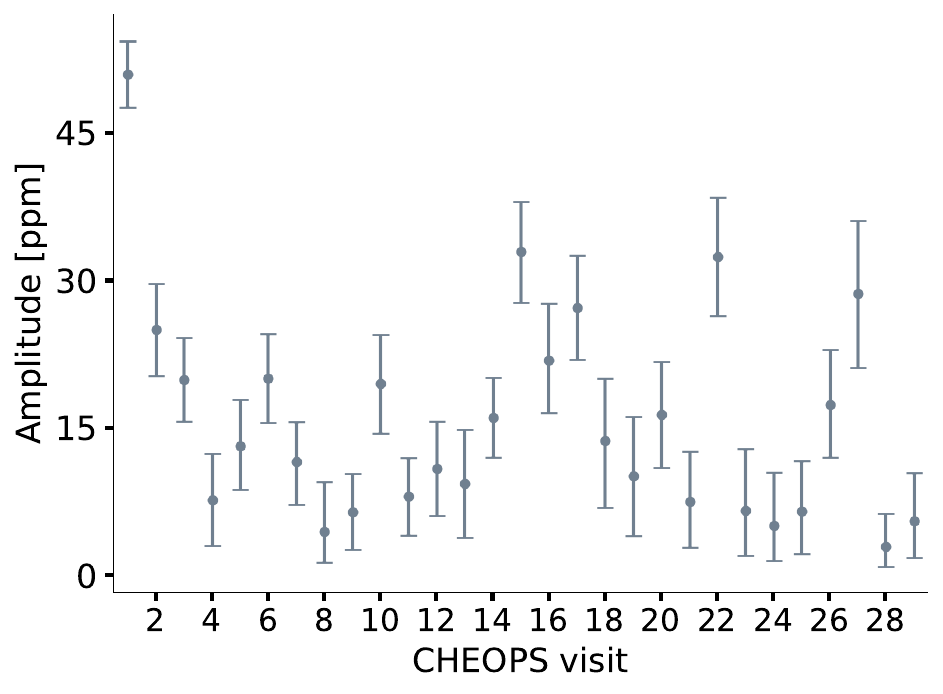}}
    \caption{Phase-curve amplitude for each CHEOPS visit. The error bars represent the 1$\sigma$ uncertainty.}
    \label{fig:amplitude}
\end{figure}

The phase-curve amplitude initially observed by \citet{Morris_2021} is seen in most of the CHEOPS visits, but the magnitude changes. The highest modulation was observed during the first visit. For visits 8, 9, 19, 21, 23, 24, 25, 28, and 29, the amplitude is consistent with zero at 2$\sigma$ significance. The difference between the lowest and highest phase-curve amplitude is 48\,ppm. The phase-curve amplitude for each visit is displayed in Fig. \ref{fig:amplitude}. A weak correlation between the phase-curve amplitude and both the transit and occultation depth is observed (see Figs. \ref{fig:transit-amplitude} and \ref{fig:occultation-amplitude}), which likely arises from the model construction. As in \citet{Morris_2021}, we do not assume that the origin of the sinusoidal signal in the flux is planetary. As a consequence, the sinusoidal signal can alter the flux level before and after a transit or occultation. 

\subsection{Phase offset}
\label{subsection:phase offset}

\begin{figure}[!t]
    \centering
    \resizebox{\hsize}{!}{\includegraphics{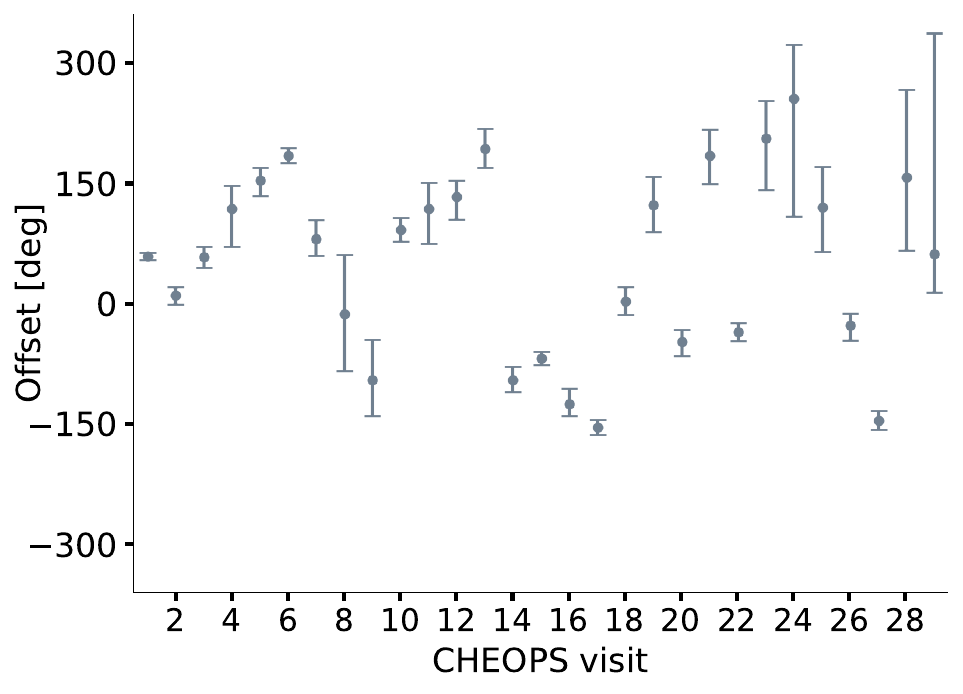}}
    \caption{Phase-curve offset for each CHEOPS visit. The error bars represent the 1$\sigma$ uncertainty.}
    \label{fig:offset}
\end{figure}

Our observations reveal a changing phase offset for the visits that varies over the complete phase parameter space. The phase offset for each visit is shown in Fig. \ref{fig:offset}. The small phase-curve amplitude and the huge uncertainty in the offset are related, as shown in Fig. \ref{fig:amplitude-offset}. If a sinusoidal function has a small amplitude, it converges in practice to a straight line, and any point is good as the peak of a sinusoid. This is the case for visits 8, 23, 24, and 28.  

The best-fit median phase offset of visits 5, 6, 12, 13, 17, 19, 21, and 25 shifts the phase peak close to the transit. This has two possible ramifications: Either the sinusoidal function, which has no physical interpretation, is a poor fit to the data, or an astrophysical event occurred during these visits that caused an excess flux during transit. If the phase offset originates at the planet, then a maximum of reflected light on the night-side would be implied if the best-fit values peaked during transit or close to it. While analysing the wide uncertainty in visits 19 and 25, we realised that the reason for is that the ingress and egress of the transit were not observed by CHEOPS. As a consequence, the mid-transit time parameter has a bimodal distribution. The same phenomenon is observed in visits 9 and 24, although in these cases, the phase peak is not close to the transit. In Appendix \ref{appendix:corner plots} we show the joint posterior distributions of a CHEOPS visit that covered the transit ingress and egress and a visit where these events could not be observed. The sinusoidal model is statistically preferred over physically motivated models (Sect. \ref{subsection:model comparison}) in most CHEOPS visits. This suggests that it is not a poor fit to the data. 

\subsection{Consecutive CHEOPS visits}
\label{subsection:consecutive visits}

\begin{figure}[h]
    \centering
    \resizebox{\hsize}{!}{\includegraphics{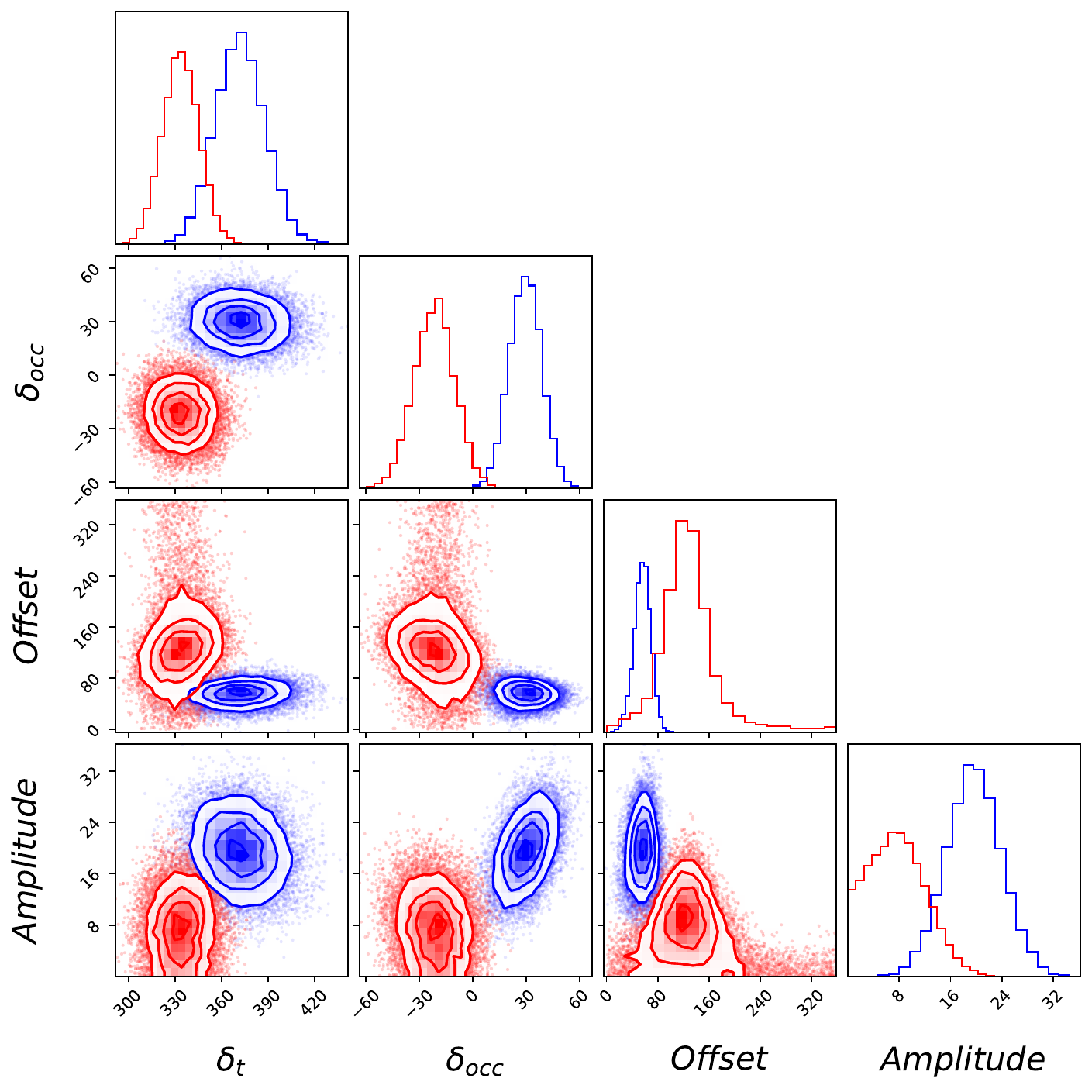}}
    \caption{Posterior distributions and joint correlations plot for visit 3 in blue and visit 4 in red. $\delta_{t}$ stands for the transit depth, $\delta_{occ}$ is the occultation depth, and the amplitude and offset refer to the phase-curve amplitude and offset, respectively.}
    \label{fig:consecutive_corner}
\end{figure}

Some visits were scheduled to start immediately after a previous observation of 55~Cnc had ended. These continuous visits provide useful information on the phase-curve change and its timescale. The consecutive visits are numbers 3 and 4, 5 and 6, 8 and 9, 10 and 11, 12 and 13, and 24 to 27. We compare consecutive visits by overplotting the posterior distributions of the transit depth, occultation depth, phase-curve amplitude, and offset to properly estimate the significance between each parameter and their joint correlations. Figure \ref{fig:consecutive_corner} shows the posterior distribution functions of phase-curve parameters and joint correlation plots of visits 3 and 4. The values of the transit depth differ at 1.9$\sigma$ between visits 3 and 4, and the significance is at 2.9$\sigma$ between visits 26 and 27. For the rest of the visit pairs, the difference is below 1.6$\sigma$. The occultation depth varies over 3$\sigma$ between visits 3 and 4. The strong variability in the occultation depth between visits 3 and 4 is shown in the top left panel in Fig. \ref{fig:consecutive_corner}, where the two posterior distributions barely overlap. The occultation depth varies significantly at 4.6$\sigma$ between visits 10 and 11 and at 3$\sigma$ between visits 24 and 25. The rest of the visit pairs show a difference below 2$\sigma$. The phase-curve amplitude differs at 1.9$\sigma$ and 2$\sigma$ between visits 3 and 4 and between visits 10 and 11, respectively. The sequential increase in amplitude from visits 25 to 27 exceeds 2.5$\sigma$ from one visit to the next. Other pairs are consistent below 1.5$\sigma$. Finally, the phase offset exhibits change over 3$\sigma$ between visits 8 and 9, 25 to 27, and between 26 and 27. Incidentally, the joint correlation plots show no significant correlation between the transit depth and phase-curve amplitude. There is low evidence of a variable transit depth and phase-curve amplitude between consecutive visits, but the occultation depth and phase offset vary significantly over 3$\sigma$ in some cases. The 2D posterior distribution correlation plot between the phase-curve amplitude and offset (third panel in Fig. \ref{fig:consecutive_corner} at the bottom from left to right) does not overlap at the 3$\sigma$ level, revealing an overall phase-curve change due to the joint change in amplitude and offset between visits 3 and 4. The same is true for visits 8 and 9, 25 and 26, and 26 and 27. Because a CHEOPS visit of 55 Cnc e lasts approximately 1.5 orbital periods, we conclude that the joint change in the parameters describing the phase curve occurs on the order of the planetary orbital timescale or approximately on the order of a day.  

\subsection{Model comparison}
\label{subsection:model comparison}

The considered models for the phase curve were compared using the leave-one-out (LOO) cross-validation \citep{Vehtari_2016}, which is a method for estimating the pointwise out-of-sample prediction accuracy from a Bayesian model. The LOO is similar to the widely applicable information criterion (WAIC), but it is more robust when the observations contain weak priors or sensitive outliers, at the cost of being computationally more expensive. It consists essentially of estimating the relative likelihood for one model to be preferred over other models in a set using the posterior samples of the MCMC. 

The top-ranked model has the lowest LOO value. The higher the difference in the LOO between models, the better the top-ranked model. In practice, the threshold of $\Delta$LOO to consider a model significantly better than others is subjective and debated \citep{McElreath_2016}, but it is common convention to consider a model significantly better if the $\Delta$LOO to the second-ranked model is greater than 10. Another relevant parameter in the model comparison is the statistical weight, which can be interpreted as an estimate of the probability that the model will make the best predictions on future data among the considered models. The values range between 0 and 1, and the sum of the weights for a set of models is equal to 1. 

\begin{figure*}[h]
    \centering
    \resizebox{\hsize}{!}{\includegraphics{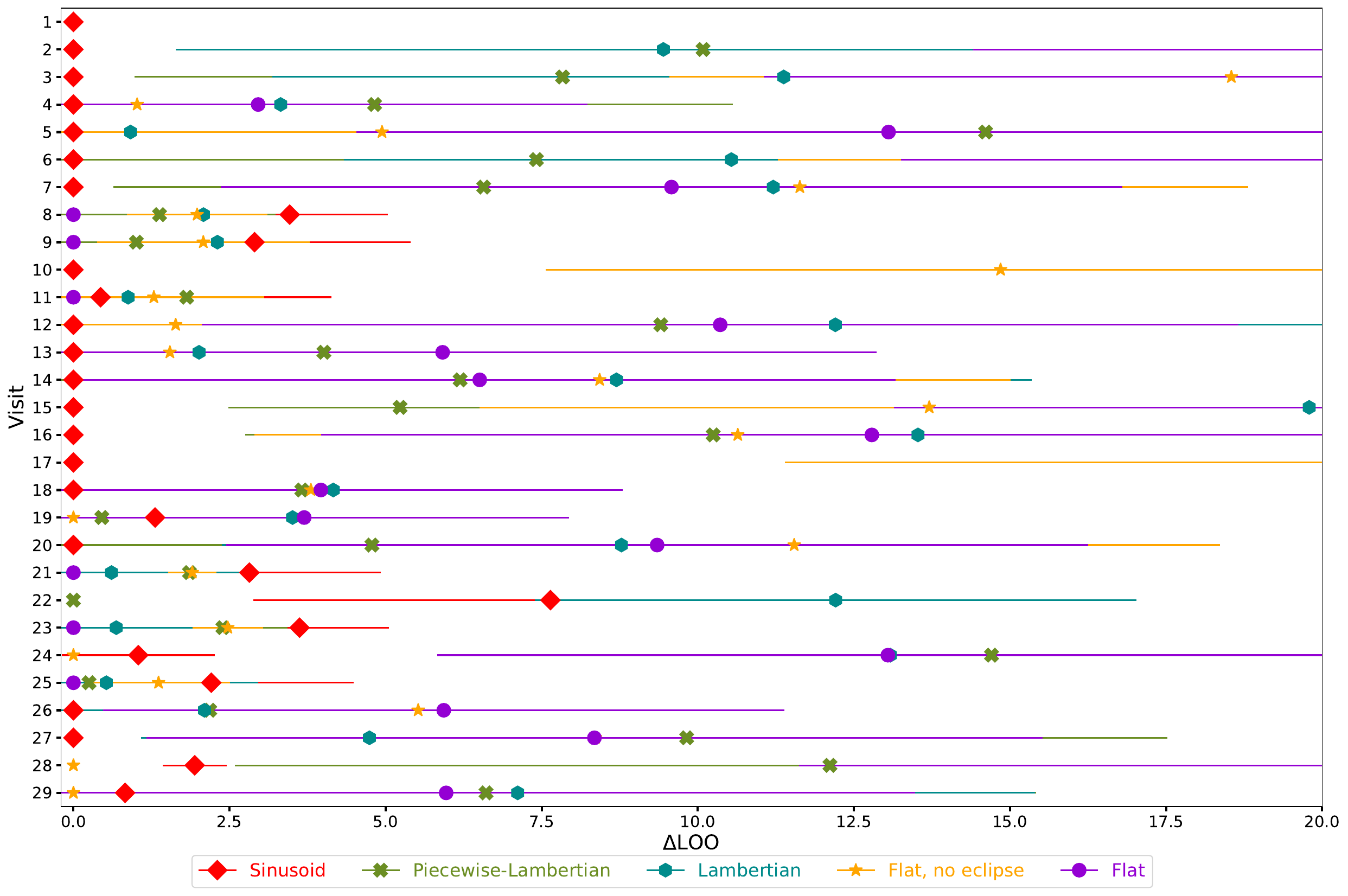}}
    \caption{Model comparison based on the LOO criterion. The best-ranked model has a $\Delta$LOO of zero. The x-axis shows the difference in the LOO for the models considered in each visit (y-axis). The error bars represent the standard error of the difference in the expected log-predictive density between each model and the top-ranked model. By definition, the standard error of the top-ranked model is zero. In visit 1, the $\Delta$LOO to the second-ranked model is above 50 and thus is not visible in the plot. The red diamond represents a sinusoidal phase curve, the green cross is the piecewise-Lambertian model, the cyan hexagon is the Lambertian sphere, the orange star is a flat baseline without an eclipse, and the violet circle is a flat baseline with the eclipse as a free parameter.}
    \label{fig:model comparison}
\end{figure*}

Because we considered five phase-curve models for a total of 29 CHEOPS visits, we present the information of the model comparison summarised in Fig. \ref{fig:model comparison}. The y-axis represents the CHEOPS visits, where each row depicts the model comparison for a specific visit. Each marker represents a phase-curve model, as indicated in the legend in Fig. \ref{fig:model comparison}. The x-axis shows the LOO relative to the the top-ranked model. Thus, the top-ranked model appears leftmost in the plot with $\Delta$LOO=0. The error bars are the standard error of the difference in the expected log-predictive density between each model and the top-ranked model. Complementary to the figure, Table \ref{tab:model comparison} reports the statistical weight of the top-ranked model.  

The simple sinusoidal modulation of the phase curve is preferred in 18 out of 29 visits, while the flat phase curve is favoured by the LOO in 6 visits. A flat phase curve without occultation is ranked best in 4 visits. The piecewise-Lambertian is preferred only in visit 22. It is worth mentioning that for the cases when a flat phase curve is preferred by the LOO, the rest of the models follow close behind. To elaborate further, in visits 8, 9, 11, 21, 23, and 25, the difference in the LOO between the best and worst model is below 5. However, the flat model takes most of the statistical weight in visits 8, 9, and 23. The best-fit results of the sinusoidal model show that the amplitude for the above-mentioned visits is consistent with zero within $2\sigma$. The small difference between the models reported by the LOO can be explained by the fact that at small amplitudes, the functions converge to a flat line. Visit 4 stands out due to its low amplitude, but it nonetheless slightly favours the sinusoidal model by the LOO. A flat phase curve without occultation is favoured in visit 28, even though the occultation depth is significant. This is even one of the deepest occultations of the dataset. While the piecewise-Lambertian is preferred only in visit 22, there is no strong indication that the observations are described best by a planet with an asymmetric albedo.

\subsection{Thermal emission}
\label{subsection:thermal emission}

We estimated the thermal contribution in the CHEOPS bandpass by retrieving a theoretical stellar spectrum from the PHOENIX stellar model \citep{Husser_2013} with an effective temperature of 5200\,K, surface gravity $\log(g)$ = 4.5 \citep{VonBraun_2011}, and a planet temperature of 2697$\pm$270\,K, which is the maximum hemisphere-averaged temperature measured by \citet{Demory_2016b} with \textit{Spitzer} observations. When the uncertainty in the brightness temperature is taken into account, the thermal contribution in the CHEOPS bandpass ranges between 3 and 11\,ppm.

\section{Discussion}
\label{section:discussion}

The CHEOPS observations of 55~Cnc~e present a puzzling case. The phase-curve amplitude and phase offset change from one visit to the next by up to 50\,ppm and span a wide range of offset angles. Consecutive visits reveal changes on the timescale of at least the orbital period.  

When we attribute the mechanism that causes this to activity of some sort, then the change between high and low phase-curve amplitude can be ascribed to periods of activity and inactivity, depending on the nature of the mechanism \citep{Sulis_2019}. A sufficiently strong grey absorber could obscure the dayside enough to produce a flat phase-curve signal and might produce flux variation due to scattering \citep{Tamburo_2018}. 

When the process behind the change in the phase-curve amplitude and offset of the signal comes from the planet or is bounded in its vicinity, then we would expect the occultation depth to be approximately twice the phase-curve amplitude. Most visits do not satisfy $2A \approx \delta_{occ}$. Additionally, the sinusoidal model infers a phase offset close to the transit or during transit for some visits. It is hard to conceal an event on the night-side of the planet to cause a stronger signal than at any other orbital phase. Thus, the origin of the variable signal is probably not at the planet. 

\subsection{Power spectrum}
\label{subsection:power spectrum simulation}

\begin{figure}
    \centering
    \resizebox{\hsize}{!}{\includegraphics{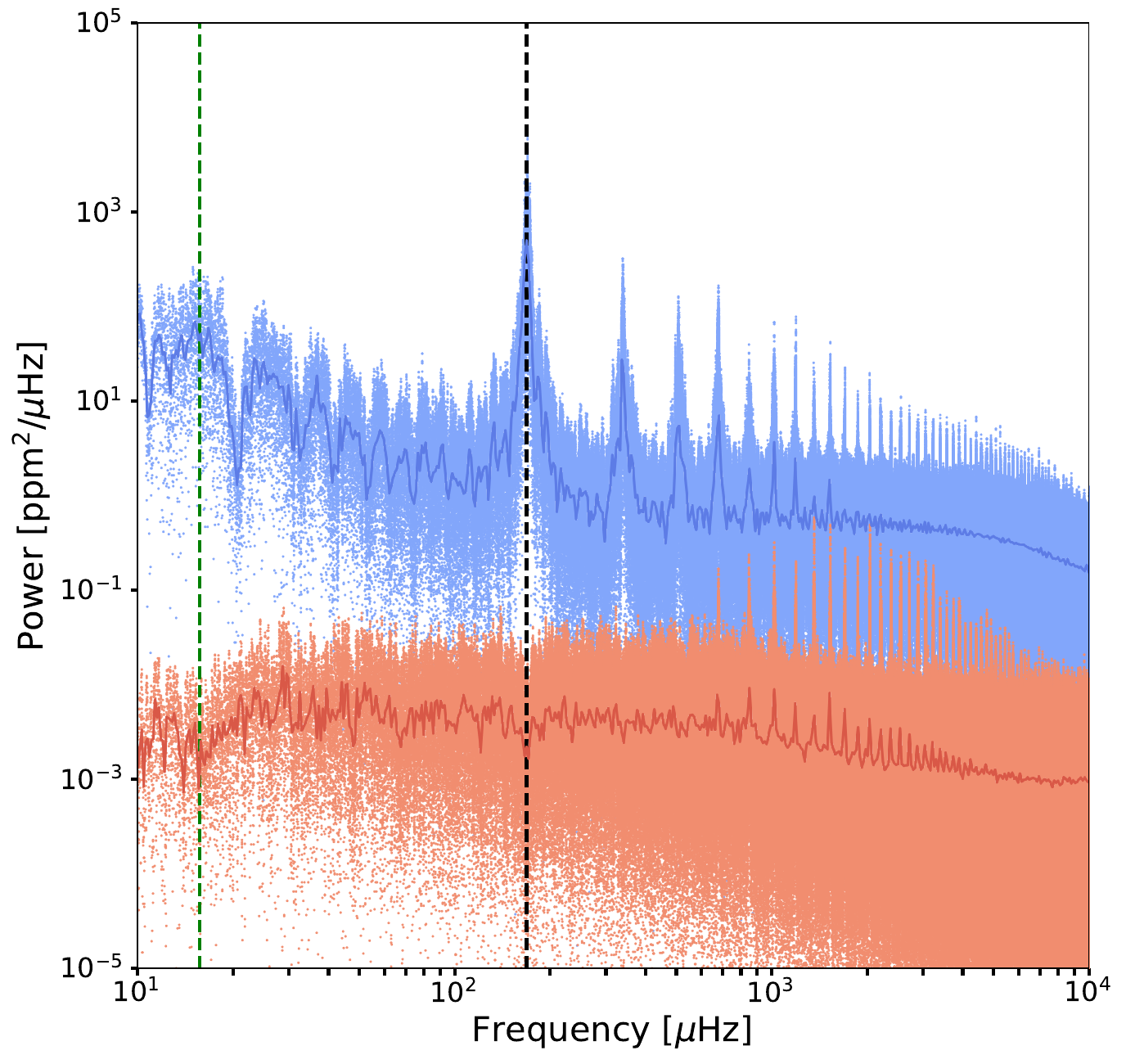}}
    \caption{Power spectrum computed with the Lomb-Scargle technique as a function of frequency. The blue dots show raw flux measurements of all CHEOPS visits, while the red dots represent the detrended flux after the phase-curve model was removed. The dark blue and dark red curves represent the binned data. The dashed green vertical line corresponds to the orbital period of 55 Cnc e, and the vertical dashed black line corresponds to the CHEOPS orbital period of 101 minutes.}
    \label{fig:Lomb-Scargle}
\end{figure}

We characterised the periodic signals in the residuals with a Lomb-Scargle periodogram (see \citet{VanderPlas_2018} for a review). In Fig. \ref{fig:Lomb-Scargle} the blue points are unprocessed flux measurements, and the darker blue curve shows the binned data. Similarly, the red points represent the power of residual data, while the darker red curve shows the binned data. The residuals consist of the CHEOPS observations after removing systematics, transit, occultation, and phase-curve model. We identified the corresponding frequency of the known periodicities such as the orbital period of planet e and the CHEOPS orbital period. Because a single CHEOPS visit of 27 hours translates into approximately 10 $\mu$Hz, we did not consider frequencies below this value. The power spectrum searches for periodicities in the time-series observations. Periodic signals can only be measured reliably for periods shorter than the visit duration \citep{Martins_2020}.

In the residuals, no strong signals remain at the orbital period of the planet. The power at the CHEOPS orbital period also shows an absence of power. However, the Lomb-Scargle periodogram in Fig. \ref{fig:Lomb-Scargle} exhibits strong peaks at higher-order harmonics of the CHEOPS orbital period at frequencies starting around 1000 $\mu$Hz. This can be indicative of an improper removal of systematic noise induced by the spacecraft. To analyse this in depth, we performed tests on simulated data.

We constructed 29 sets, each one consisting of 2000 points with an observing efficiency of 56\% (see Table \ref{tab:obslog}), lasting for 1.5 orbital periods. The gaps of 100 minutes in the dataset simulate the CHEOPS occultation caused by Earth. Between two consecutive sets, which represent CHEOPS visits, we added gaps of random durations. The points were randomly drawn from a normal distribution representing white noise on a level comparable to the real CHEOPS observations. In addition, we added correlated noise with a 1D Gaussian filter with a period that matched the stellar rotation period of 38.8 days \citep{Bourrier_2018}. The resulting power spectra show peaks at a frequency corresponding to the 100-minute duration of the gaps due to the CHEOPS orbital configuration and its higher harmonics. The peaks in our residuals might therefore be explained by the CHEOPS orbital configuration and do not necessarily imply an insufficient systematics detrending. Moreover, the absence of power in the periodogram at the CHEOPS orbital period shows that there are no strong signals at this frequency in the residuals. 

Another source of time-correlated astrophysical noise that we expect in the power spectrum is stellar granulation of the host 55 Cnc, which occurs at multiple time  and length scales \citep{Rast_2003}. A  granulation phenomenon called supergranulation has a characteristic timescale similar to the orbital period of 55 Cnc e, and this could give rise to the apparent phase variations that change in amplitude and phase over time. It is difficult to reliably measure the excess power on the supergranulation timescale with these CHEOPS observations because the duration of each visit in this work is about one granulation timescale. Future observations with longer visit durations, or theoretical advances in modeling the long-timescale granulation phenomenon, can address this hypothesis.

\subsection{Phase-curve amplitude variability}
\label{subsection:amplitude periodicity}

\begin{figure}
    \centering
    \resizebox{\hsize}{!}{\includegraphics{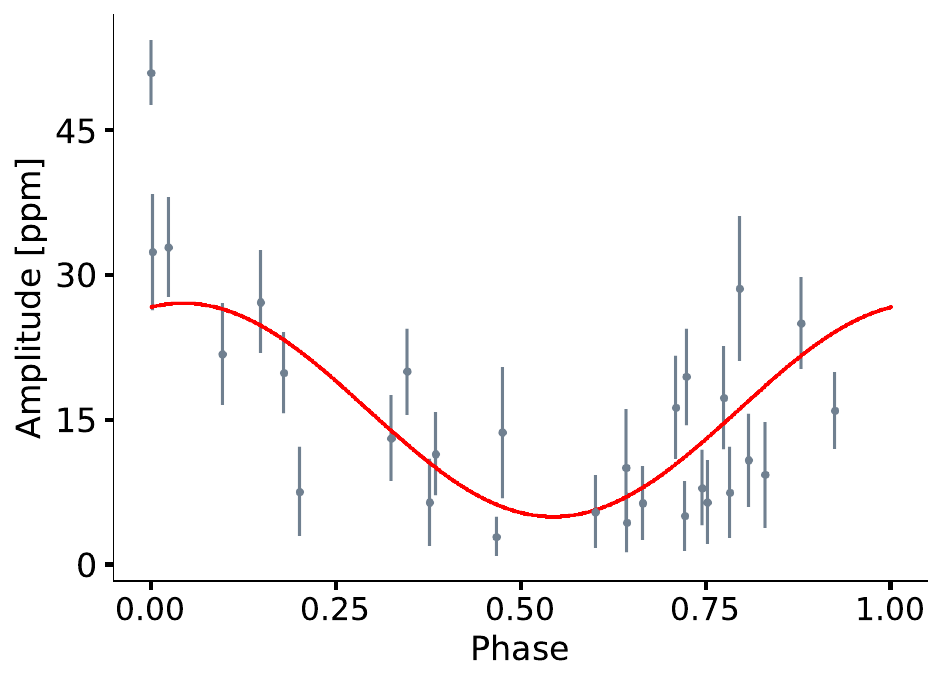}}
    \caption{Phase-curve amplitude of every CHEOPS visit phase-folded at the best-fit period of 51.9 days. The sinusoidal function is overplotted in red.}
    \label{fig:amplitude phase-folded}
\end{figure}

Motivated by the 47.3-day periodicity in the occultation depth of 55~Cnc~e found by \citet{Demory_2022}, we investigated the putative variability timescale of the phase-curve amplitude. Similar to \citet{Tamburo_2018} and \citet{Demory_2022}, we constructed three different models to reproduce the amplitude as a function of time. The first model consists of a flat line. The second model is a linear function, where the free parameters are the slope and intercept. The last model is a sine function of the form $A\sin(2\pi t/P+B)+C$, where $t$ is BJD time, and $P$ is the variability period. The caveat of this analysis is that there is no precise timing associated with the amplitude of the phase, in contrast to the precise mid-eclipse time used in \citet{Demory_2022}. As an approximation, we used the mean BJD time of each CHEOPS visit. The periodogram of the amplitude of each CHEOPS visit shows a maximum peak at 52.25 days. This peak has no dominant power compared to other peaks, and thus, we used the maximum value only as a reference to set a uniform prior between 40 and 70 days. We fitted our models in an MCMC routine. The sine function is favoured by the LOO because it carries all the statistical weight, and with a $\Delta$LOO=11 to the linear function, ranked second. The flat line is ranked last, with $\Delta$LOO=14. We find a period of $51.9 \pm 1.4$ days, $A=11.1 \pm 2.3$\,ppm, $B=73.85 \pm 21.84$ degrees, and $C=16.2 \pm 1.6$\,ppm. The amplitudes of each visit are phase-folded at the best-fit period and are shown in Fig. \ref{fig:amplitude phase-folded}. From the best-fit sinusoidal model, we infer an estimate reference timing of the local maximum at 2458934.98 BJD. 

The period is similar to the 47.3-day periodicity on the occultation depth \citep{Demory_2022}. 
At the present time, it is unclear whether the periodicity in the occultation depth in \citet{Demory_2022} and in the phase-curve amplitude are related. 
The 51.9-day period is absent from the power spectrum of the CHEOPS residuals (Fig. \ref{fig:Lomb-Scargle}). The rotation period of the star of 38.8 days from combined photometry and spectroscopy \citep{Bourrier_2018} and 42.7 from photometry \citep{Fisher_2008} appear close to the 51.9-day period. We also note that the orbital period of the non-transiting planet c is 44.4 days. An approximate computation of the reflected light that planet c could contribute yields a maximum value of 2.7 ppm. For this estimate, we assumed no thermal emission and a geometric albedo of unity to obtain an upper limit. Because planet c is not transiting and its inclination is unknown, RV measurements \citep{Bourrier_2018} estimate a mass comparable to the mass of Saturn, and thus, we used the nominal radius of Saturn for the calculation. The estimated reflected light that could leak into the field of view of CHEOPS is too low to match the 11 ppm amplitude found in the periodic signal of the phase-curve amplitude, and it is thus unlikely to be the sole cause for the signal.
The origin of our 51.9-day signal is unknown. 

\subsection{Observed flux dips}
\label{subsection:dips}

Visual inspection of the detrended flux and bins in Fig. \ref{fig:gallery} reveals a decrease in flux, so-called dips, outside of the transit and occultation of 55~Cnc~e. The most conspicuous dip is found in visit 17 after the second occultation, approximately at BJD time 2459251.98. Other identified dips occur during visits 26, 27, and 28 at BJD time 2459595.3, 2459595.95, and 2459630.28, respectively. In Table \ref{tab:dips} we summarise the identified dips with an estimated time of the event and depth. 

Another interesting case is visit 10, where the residuals in Fig. \ref{fig:gallery residuals 1} show three small dips that occur roughly every 0.5 days at approximately BJD time 2459229.1, 2459229.59, and 2459230. However, inspection of the power spectrum on the complete dataset of the residuals does not reveal a significant periodicity at 0.5 days (the corresponding frequency of 23.1 $\mu$Hz). 

\begin{table}[h]
        \centering\setstretch{1.5}
        \caption{Identified flux drops in the observations and their estimated depth.}
        \begin{tabular}{c c}
            \hline
            \hline
            Time [BJD] & Depth [ppm] \\
            \hline 
            2459229.1 & 20 \\
            2459229.59 & 50 \\
            2459230 & 10 \\
            2459251.98 & 80 \\
            2459595.3 & 10 \\
            2459595.95 & 30 \\
            2459630.28 & 50 \\
            \hline
        \end{tabular}
        \label{tab:dips}
\end{table}

We checked whether another planet in addition to planet e might be transiting during the CHEOPS visits. 55~Cnc~A is known to host five planets, of which planet e is the only transiting planet.  We used the ephemeris of the remaining planets in Table 3 of \citet{Bourrier_2018} to compute predicted transit times. Only during visit 4 did a predicted possible transit of planet b coincide with the observing time. Phase-folding of the observations provided no hint of a transit, and running an MCMC searching for the planet was negative. The dips are not related to a transit or occultation of planets b, c, d, or f. After carefully inspecting the dips and each frame of the corresponding visit, we note that the dip durations coincide exactly with the duration of a CHEOPS orbit. We therefore suspect that they are systematic.

Based on the analysis of the observations on the phase-curve, we discuss possible mechanisms that might have caused the phase-curve amplitude and shift variability. Previous research suggested that refractory material \citep{Tamburo_2018} or an inhomogeneous circumstellar dust torus \citep{Demory_2015, Sulis_2019, Morris_2021} might explain the observations, as might volcanic activity \citep{Demory_2015} or star-planet interaction \citep{Demory_2015, Sulis_2019, Folsom_2020, Morris_2021}.

\subsection{Dust in the environment of 55~Cnc~e}
\label{subsection:dust}

In the following sections, we build a toy model to study dust in the orbital environment of 55~Cnc~e. First, we estimate the required material to obscure the stellar light by an arbitrary fraction. Then we constrain the composition of the material by estimating its characteristic sublimation timescale and comparing this to the variability timescales in the data. Finally, we compute the motion of the dust in the system after it escapes the planet to determine whether the ejected material could form a circumstellar torus. 

Because of the uncertainty on the composition and interior of the planet, we consider characteristic species for a rocky planet and lava worlds: silicon monoxide, fayalite (iron-rich end-member\footnote{An end-member is a mineral at the extreme end of a mineral series in terms of purity, often described as solid solutions with varying compositions of some chemical elements.} of olivine), enstatite (an end-member of pyroxene), forsterite (magnesium-rich end-member of olivine), $\alpha$-quartz and amorphous quartz, corundum, silicon carbide, and graphite. The silicates pyroxene and olivine are expected in rocky exoplanets because they predominate in Earth's mantle \citep{Perez-Becker_2013}. $\alpha$-quartz refers to quartz with a trigonal structure, which is a crystalline unit that looks like an oblique cube whose angles in the corners are equal, but not rectangular. Table \ref{tab:dust species} lists the selected species.

\subsubsection{Estimated mass loss}
\label{subsubsection:ejected mass}

The amplitude change in the phase curve reveals a maximum difference of approximately 50\,ppm, as shown in Fig. \ref{fig:amplitude}. We investigated the estimated amount of material required to produce a 50\,ppm change in flux. For the purpose of this computation, we assumed instead of considering material causing variation out of transit that the material composed of grey absorbing grains transited the star. We determined the material that is required to obscure the star by  50\,ppm. 

Following \citet{Perez-Becker_2013}, we first computed the total mass required to absorb or scatter a certain fraction $f$ of the starlight. We assumed that the dust is optically thin (otherwise, the surface of the planet would cool down enough so that the material would not be produced in the first place \citep{Rappaport_2012}). We let the dust cover an area $A$ of the stellar disk $\pi R_{\star}^{2}$ and have an optical depth $\tau_{d}=\overline{m}_{d}\kappa_{d}$, where the subscript $d$ represents the dust, $\overline{m}_{d}$ is the grain mass per unit area, and $\kappa_{d}$ is the opacity. Then the fraction of starlight is given by
\begin{IEEEeqnarray*}{rll}
\label{eqn:fraction of starlight}
    &f& = \frac{A}{\pi R_{\star}^{2}}\overline{m}_{d}\kappa_{d} \IEEEyesnumber
.\end{IEEEeqnarray*}

When we consider the grains to be spherical with radii $s$ and internal density $\rho_{int}$, the opacity is given as 
\begin{IEEEeqnarray*}{rll}
\label{eqn:opacity}
    &\kappa_{d}& = \frac{3}{4\rho_{int}s } \IEEEyesnumber
.\end{IEEEeqnarray*}

The total mass in dust covering the star is given as a function of the grain mass per unit area $\overline{m}_{d}$ and total covered area $A$ as
\begin{IEEEeqnarray*}{rll}
\label{eqn:mass grain}
    &M_{d}& = \overline{m}_{d}A \IEEEyesnumber
.\end{IEEEeqnarray*}

Substituting Eq. (\ref{eqn:opacity}) into Eq. (\ref{eqn:fraction of starlight}), solving for $\overline{m}_{d}A,$ and substituting into Eq. (\ref{eqn:mass grain}), we obtain a total mass of 
\begin{IEEEeqnarray*}{rll}
\label{eqn:mass grain 2}
    &M_{d}& = \frac{4\pi}{3}f\rho_{int}R_{\star}^{2}s \IEEEyesnumber
.\end{IEEEeqnarray*}

To obtain a numerical estimate, we first took possible grain sizes into account that are opaque in the CHEOPS bandpass and transparent in the \textit{Spitzer} bandpass \citep{Morris_2021}, resulting in a plausible range of particle radii $0.1 \le s \le 0.7 \ \mu$m. The grain density was chosen to be 3\,g\,cm$^{-3}$, similar to the density of forsterite and enstatite. The radius of the star was $R_{\star}=0.94\,R_{\odot}$. For a fraction of obscured starlight of $f=50$\,ppm, we estimate a mass between $2.7\times 10^{9}$ and $1.88 \times 10^{11}$\,kg. Finally, the mass-loss rate was estimated by dividing by the planetary orbital period of 0.74 days, yielding a rate $\dot{M}_{d}$ between $1.33\times 10^{12}$ and $9.32 \times 10^{13}$ kg/yr.

To place this value in perspective, we compared it to the mass loss due to photoevaporation of the planetary atmosphere driven by the combined X-ray and extreme ultra-violet (EUV) flux from the host star, abbreviated as XUV. In this case, the mass-loss rate due to photoevaporation is \citep{Rappaport_2014}
\begin{IEEEeqnarray*}{rll}
\label{eqn:photoevaporation}
    &\dot{M}_{evap}& = \frac{3F_{XUV}}{4G\rho_{p}} \IEEEyesnumber
,\end{IEEEeqnarray*}
where $F_{XUV}$ is the XUV flux from the host star at the planet, $\rho_{p}$ is the bulk density of the planet, and $G$ is the gravitational constant. The estimate was made without an assumption on the planetary atmosphere. \citet{Sanz-Forcada_2011} collected X-ray luminosities for 82 stars and inferred EUV luminosities using coronal models. The star 55~Cnc~A has available X-ray observations. The XUV flux at 55~Cnc~e is 870.96 erg s$^{-1}$cm$^{-2}$ (see Table 6 of \citet{Sanz-Forcada_2011}). Using appropriate parameters for the bulk density of 55 Cnc e \citep{Bourrier_2018}, we obtain a mass-flux rate due to photoevaporation of $4.64 \times 10^{13}$ kg/yr.

Our estimated mass loss due to the amount of material obscuring the star is comparable to the mass-loss rate due to photoevaporation, which suggests that the amount of material required to cause a 50\,ppm change is plausible. However, photoevaporation is not the only mechanism leading to atmospheric escape. Tidally driven volcanism, as well as Jeans escape \citep{Oza_2019} and plasma-driven atmospheric sputtering feeding a plasma torus, are mass-loss processes that could occur on a super-Earth \citep{Gebek_2020}. The main challenge for the required mass to escape the planet is the high escape velocity of approximately 24 km/s. However, violent plumes such as those observed in Io \citep{Jessup_2012} could provide the material to the circumstellar environment. The plume speeds in Io are approximately 0.5 km/s \citep{Lellouch_1996}, which is lower than the Io escape velocity of 2.6 km/s. Even so, plume material escapes the atmosphere with the aid of additional mechanisms such as sputtering by charged particles \citep{Geissler_2007}. A complex model combining different mass-loss processes is required to determine whether the amount of material can reach the escape velocity. 

\subsubsection{Characteristic sublimation timescales}
\label{subsubsection:sublimation timescales}

Due to the close-in orbit of 55 Cnc e and its likely tidally locked configuration, the silicates vaporise at the day-side surface temperature of $\sim$2700\,K \citep{Demory_2016b}  and form an atmosphere whose equilibrium vapour pressure is given by the Clausius-Clapeyron equation,
\begin{IEEEeqnarray*}{rll}
\label{eqn:vapour pressure}
    &P_{\rm vapour}(T)& = \exp\left(-\frac{\mu m_{H}L_{sub}}{k_{B}T}+B\right) \IEEEyesnumber
,\end{IEEEeqnarray*}
where $\mu$ is the molecular mass of gas released from dust due to sublimation, $m_{H}$ is the atomic mass unit \citep{Kimura_2002}, $L_{sub}$ is the latent heat of sublimation, $k_{B}$ is Boltzmann's constant, and $B$ is a constant composition-dependent sublimation parameter. We express Eq. (\ref{eqn:vapour pressure}) as
\begin{IEEEeqnarray*}{rll}
\label{eqn:vapour pressure 2}
    &P_{\rm vapour}(T)& = \exp\left(-\frac{\mathcal{C}}{T}+B\right) \IEEEyesnumber
,\end{IEEEeqnarray*}
with 
\begin{IEEEeqnarray*}{rll}
\label{eqn:vapour pressure 3}
    &\mathcal{C}& = \frac{\mu m_{H}L_{sub}}{k_{B}} \IEEEyesnumber
.\end{IEEEeqnarray*}

The sublimation parameters for the selected species are shown in Table \ref{tab:dust species}. 

\begin{table*}[h]
        \centering\setstretch{1.5}
        \caption{Sublimation parameters for the selected dust species \citep{van-Lieshout_2014}.}
        \begin{tabular}{c c c c c c}
            \hline
            \hline
            Dust species & $\mu$ & $\rho$\tablefootmark{a} [g/cm$^{3}$] & $\mathcal{C}$ [K] & B & $L_{sub}$ [J/kg] \\
            \hline 
             Enstatite (MgSiO$_{3}$) & 100.389 & 3.20 & 68 908 & 38.1 & 5 707 129.2 \\
             
             Forsterite (Mg$_{2}$SiO$_{4}$) & 140.694 & 3.27 & 65 308 & 34.1 & 3 859 446.2 \\
            
             Silicon monoxide (SiO) & 44.085 & 2.13 & 49 520 & 32.5 & 9 339 507.5 \\
             
             Fayalite (Fe$_{2}$SiO$_{4}$) & 203.774 & 4.39 & 60 377 & 37.7 & 2 463 524.8 \\
             
             Quartz (SiO$_{2}$) & 60.084 & 2.6 & 69 444 & 33.1 & 9 609 605.4 \\
             
             Corundum (Al$_{2}$O$_{3}$) & 101.961 & 4.00 & 77 365 & 39.3 & 6 308 769 \\
             
             Silicon carbide (SiC) & 40.10 & 3.22 & 78 462 & 37.8 & 16 268 563 \\
             
             Graphite (C) & 12.011 & 2.16 & 93 646 & 36.7 & 64 825 257 \\
            \hline
        \end{tabular}
        \label{tab:dust species}
        \tablefoot{
        $\mu$ is the molecular weight, $\rho$ is the density of the dust, $\mathcal{C}$ and B are sublimation parameters, and $L_{sub}$ is the latent heat of sublimation.
        \tablefoottext{a}{Density values are listed in Table 5 of \citet{van-Lieshout_2016}.}
        }
\end{table*}

The mass-loss rate of dust particles due to sublimation is given by \citep{Kimura_2002}
\begin{IEEEeqnarray*}{rll}
\label{eqn:mass-loss sublimation}
    &\frac{dm}{dt}& = -S \sqrt{\frac{\mu m_{H}}{2\pi k_{B}T}}P_{\rm vapour}(T) \IEEEyesnumber
,\end{IEEEeqnarray*}
where $S$ is the surface area of the aggregate particle. We assumed that a dust grain is composed of $N$ identical spheres with radius $s$, so that $S=4 \pi Ns^{2}$.

We calculated the equilibrium temperature $T$ of the dust in local thermodynamic equilibrium through an energy balance between the absorption rate of stellar radiation and the thermal emission and energy loss through sublimation \citep{Gail_2013},
\begin{IEEEeqnarray*}{rll}
\label{eqn:energy balance}
    \Omega \int C_{abs}(\mathit{n}, \mathit{x})B_{\star}(\lambda)d\lambda \ &=& \ 4 \pi \int C_{abs}(\mathit{n}, \mathit{x})B_{\lambda}(\lambda , T)d\lambda  -\frac{dm}{dt}L, \\ \IEEEyesnumber
\end{IEEEeqnarray*}
where $B_{\star}(\lambda)$ is the solar radiance, and $B_{\lambda}(\lambda, T)$ is the Planck function of the dust \citep{Kimura_2002}. The solid angle subtended by the star at a distance $r$ is given by 
\begin{IEEEeqnarray*}{rll}
\label{eqn:solid angle}
    \Omega &=& \ 2 \pi\left[1-\sqrt{1-\left(\frac{R_{\star}}{r}\right)^{2}} \right] \IEEEyesnumber
.\end{IEEEeqnarray*}

The absorption cross sections $C_{abs}$ in Eq. (\ref{eqn:energy balance}) depend on the complex refractive index $\mathit{n}$, the size parameter $\mathit{x}=2 \pi s / \lambda,$ and the structure of the particle. These cross sections were computed using the Mie theory. We used the program \texttt{LX-MIE} \citep{Kitzmann_2018} to retrieve the absorption cross sections for each species listed in Table \ref{tab:dust species} for dust radii 0.1, 0.2, 0.3, 0.4, 0.5, 0.6, and 0.7 $\mu$m and for a wide discrete range of wavelength values available in the \texttt{LX-MIE} collection. This narrow range of particle sizes satisfies the conditions of being opaque in the optical \citep{Morris_2021}, but transparent in the IR \citep{Demory_2015}. The dataset of \texttt{LX-MIE} contains optical properties of 32 condensates (see Table 1 of \citet{Kitzmann_2018} for more details). The equilibrium temperature for each grain size was pretabulated according to Eq. (\ref{eqn:energy balance}).   

Finally, after the temperature of the grains reached a state of equilibrium, the characteristic timescales for sublimation were estimated. For this purpose, we considered the grain mass as a function of its density and volumen of spheres. Taking the first derivative, we obtain

\begin{IEEEeqnarray*}{rll}
\label{eqn:mass-loss sublimation 2}
    \frac{dm}{dt} \ &=& \ 4\rho \pi N s^{2} \frac{ds}{dt} \IEEEyesnumber
,\end{IEEEeqnarray*}
and rewriting Eq. (\ref{eqn:mass-loss sublimation}) results in
\begin{IEEEeqnarray*}{rll}
\label{eqn:mass-loss sublimation 3}
    \frac{ds}{dt} \ &=& \ -\sqrt{\frac{\mu m_{H}}{2\pi k_{B}T}}\frac{P_{\rm vapour}(T)}{N\rho} \IEEEyesnumber
.\end{IEEEeqnarray*}
The advantage of Eq. (\ref{eqn:mass-loss sublimation 3}) is that it relates the grain radii and time explicitly. The right-hand side of the equation only depends on the temperature and properties of the grain. After tabulating all relevant values, we solved the differential equation numerically to estimate the required time for a dust grain of a given size to sublimate. We integrated numerically using a trapezoidal method for an initial grain radius until it reached a threshold value of $s=0.001 \ \mu$m.

\begin{table*}[h]
        \centering\setstretch{1.5}
        \caption{Characteristic sublimation timescales in seconds of a single dust grain of the selected species with radii between 0.1 and 0.7 $\mu$m.}
        \begin{tabular}{c c c c c c c c}
            \hline
            \hline
        \multirow{2}{*}{Dust species} & \multicolumn{7}{c}{Grain size [$\mu$m]} \\ 
        \cmidrule{2-8} 
        \vspace{-2mm}
            & 0.1 & 0.2 & 0.3 & 0.4 & 0.5 & 0.6 & 0.7 \\ 
        \multirow{2}{*}{} & \multicolumn{7}{c}{Sublimation timescale [s]} \\
         \hline
             Enstatite (MgSiO$_{3}$) & 15.18 & 16.31 & 16.90 & 17.32 & 17.65 & 17.92 & 18.16 \\
             
             Glassy enstatite (MgSiO$_{3}$) & 11.88 & 12.87 & 13.42 & 13.81 & 14.12 & 14.36 & 14.57 \\
             
             Forsterite (Mg$_{2}$SiO$_{4}$) & 10.39 & 11.18 & 11.59 & 11.89 & 12.12 & 12.33 & 12.51  \\
            
             Silicon monoxide (SiO) & 0.37 & 0.38 & 0.39 & 0.40 & 0.41 & 0.42 & 0.43 \\
             
             Fayalite (Fe$_{2}$SiO$_{4}$) & 3.66 & 3.84 & 3.91 & 3.96 & 4.00 & 4.03 & 4.06 \\
             
             $\alpha$-Quartz (SiO$_{2}$) & 10755.82 & 20111.49 & 27817.78 & 33544.02 & 37167.47 & 39041.08 & 39820.98 \\
             
             Amorphous quartz (SiO$_{2}$) & 2066.63 & 3707.10 & 5106.53 & 6241.22 & 7069.07 & 7591.62 & 7874.51 \\
             
             Corundum (Al$_{2}$O$_{3}$) & 0.63 & 0.66 & 0.67 & 0.69 & 0.70 & 0.71 & 0.72 \\
             
             Silicon carbide (SiC) & 16152.55 & 16153.81 & 16157.70 & 16206.08 & 16530.99 & 18522.19 & 26088.48 \\
             
             Graphite (C) & 4.43 & 421.89 & 5886.80 & 25859.17 & 63888.80 & 113996.63 & 168262.97 \\
            \hline
        \end{tabular}
        \label{tab:sub timescales}
\end{table*}

For a single sphere ($N$=1), the characteristic sublimation timescale of the selected species is shown in Table \ref{tab:sub timescales}. Silicates made of pyroxene, olivine, and fayalite survive less than a minute. Silicon monoxide and corundum have an even shorter sublimation timescale on the order of a second. An $\alpha$-quartz grain remains in the environment for up to approximately 11 hours before sublimating, while for amorphous quartz, the sublimation time is around two hours. Silicon carbide has a survival time between 4 and 7 hours. The range of characteristic sublimation timescales of graphite spans a wide range of values for different grain radii, translating into sublimation times of between 4 seconds and up to 46 hours. Because the phase variation of 55~Cnc~e is observed at least on the orbital timescale of around 17.7 hours, most of the species for the radius range sublimate long before they would be able to produce the phase-curve variability. Only graphite and $\alpha$-quartz between 0.3 and 0.7 $\mu$m, as well as a 0.7 $\mu$m grain of silicon carbide, survive over multiple hours. Ultimately, larger grains composed of graphite and $\alpha$-quartz have a sublimation time comparable to the orbital period of the planet. For an Earth-like mantle, the short sublimation lifetimes of pyroxene and olivine would require the planet to continuously supply material to the atmosphere to produce variability on the timescale of hours. Such a continuous replenishment of material to the circumstellar environment would result in comet-like tail shape in the transit, which has not been observed. Because the composition of the outgassed material is connected to the composition of the planetary mantle \citep{Gaillard_2014}, it seems unlikely that these elements originate from the planet and reach the circumstellar environment or at least are abundant enough to cause the observed variability.

\subsubsection{Stellar radiation pressure}
\label{subsubsection:radiation pressure}

We considered the scenario of material that is ejected from the planet through, for example, explosive volcanism or high-altitude condensation \citep{Mahapatra_2017}. As soon as the dust leaves the planet, it will be subject to radiation pressure by the star. The ratio of radiation pressure to gravity acting on a dust particle with mass $m_{d}$ is given by \citep{Kimura_2002}\footnote{The correct formulation of the solid angle used here to calculate the ratio of radiation pressure to gravity, $\Omega=\pi r^{2}/R_{\star}^{2}$, can be found in \citet{Kimura_2004}.}
\begin{IEEEeqnarray*}{rll}
\label{eqn:beta}
    \beta \ &=& \ \frac{\pi R_{\star}^{2}}{GM_{\star}m_{d}c} \int B_{\star}(\lambda)C_{pr}(\mathit{n}, \mathit{x})d\lambda  \IEEEyesnumber
,\end{IEEEeqnarray*}
where $M_{\star}$ is the mass of the host star, $c$ is the speed of light, and $C_{pr}(\mathit{n}, \mathit{x})$ is the radiation pressure cross section, defined as $C_{pr}=C_{abs}+(1-g_{0})C_{sca}$, with $C_{abs}$ and $C_{sca}$ the absorption and scattering cross section, respectively; and $g_{0}$ is the scattering asymmetry parameter. The $g_{0}$ parameter describes how isotropic or anisotropic the scattering is. It is usually tabulated as a function of the wavelength and particle size. For $g_{0}=0,$ light is scattered equally in all directions\footnote{An asymmetry parameter of $g_{0}=0$ does not necessarily imply isotropic scattering \citep{Kitzmann_2018}. To be precise, it corresponds to a symmetric phase function \citep{Heng_2017}.}. A positive value of the asymmetric parameter indicates forward scattering, while for a negative value, backscattering prevails. We computed these quantities with \texttt{LX-MIE}. By examining the output of \texttt{LX-MIE}, the asymmetry parameter in the CHEOPS bandpass for particle sizes between 0.1 and 0.7 $\mu$m shows that forward scattering dominates for all considered species in the shorter wavelengths. However, forward scattering is not dominant for the small grain sizes. For the smallest grains, the asymmetry parameter tends asymptotically to a symmetric phase function for longer wavelengths in the CHEOPS range.

A ratio of radiation pressure to gravity above 0.5 leads to an unbounded trajectory, while values below this threshold correspond to closed orbits. In the presence of radiation pressure and gravity, we can interpret the two opposing forces as an effective reduced gravitational field compared to the planet and defined as $g_{eff}=GM_{\star}(1-\beta)/r^{2}$. Material in bounded orbits moves in a Keplerian ellipse with the periastron at the location where it was released. Conservation of energy and angular momentum between the grain and the planet provide information regarding the eccentricity and semi-major axis of the grain \citep{Rappaport_2014},

\begin{IEEEeqnarray*}{rll}
\label{eqn:eccentricity and semimajor axis}
    e_{d} \ &=& \frac{\beta}{1-\beta}; \ \ \ a_{d} = a_{p}\frac{1-\beta}{1-2\beta}  \IEEEyesnumber
,\end{IEEEeqnarray*}
where $a_{p}$ is the semi-major axis of the planet. With Kepler's third law, we then computed the period of a grain released by the planet and subject to radiation pressure. 

\begin{figure}
    \centering
    \resizebox{\hsize}{!}{\includegraphics{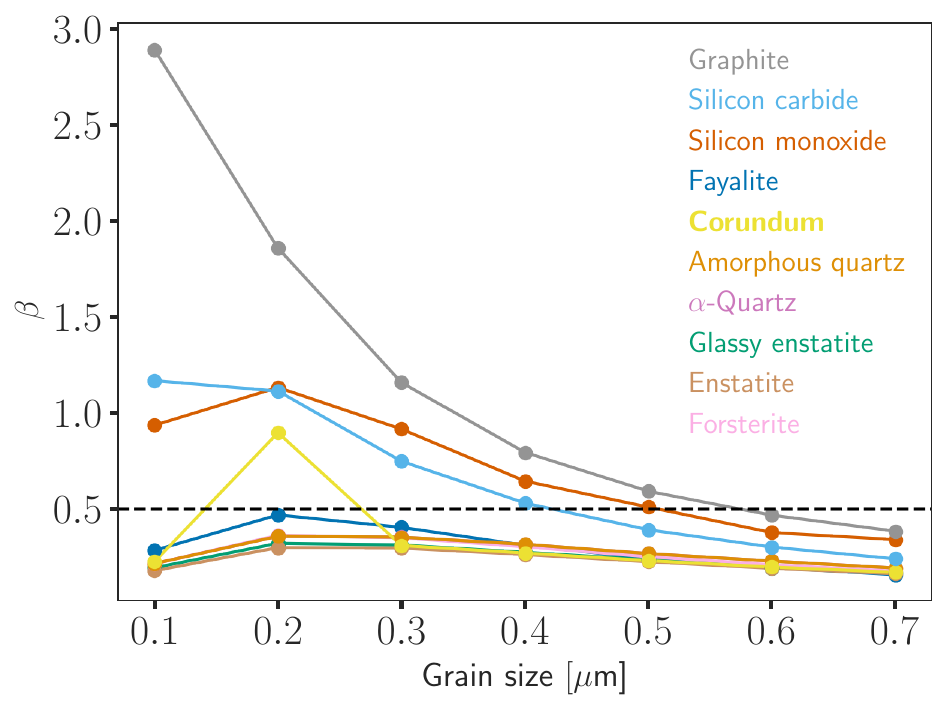}}
    \caption{Ratio of radiation pressure to gravity $\beta$ as a function of grain radius. The values below the horizontal dashed black line ($\beta$=0.5) lead to closed orbits, while the values above the line lead to unbounded motion.}
    \label{fig:betas}
\end{figure}

We used Eq. (\ref{eqn:beta}) to estimate the ratio of radiation pressure to gravity for our range of grain radii and compositions. In Fig. \ref{fig:betas} the ratio of radiation pressure to gravity for each considered species is shown. Grains with a ratio above 0.5 will be blown away, while grains below this ratio will orbit the star. The latter case is of particular interest because of the possibility of forming a circumstellar torus of dust. Most silicates within our size range move in a closed orbit, while graphite and silicon carbide grains smaller than 0.5 $\mu$m experience a radiation pressure that is too strong and are blown away. The same occurs to silicon carbide smaller than 0.4 $\mu$m.

\begin{figure}
    \centering
    \resizebox{\hsize}{!}{\includegraphics{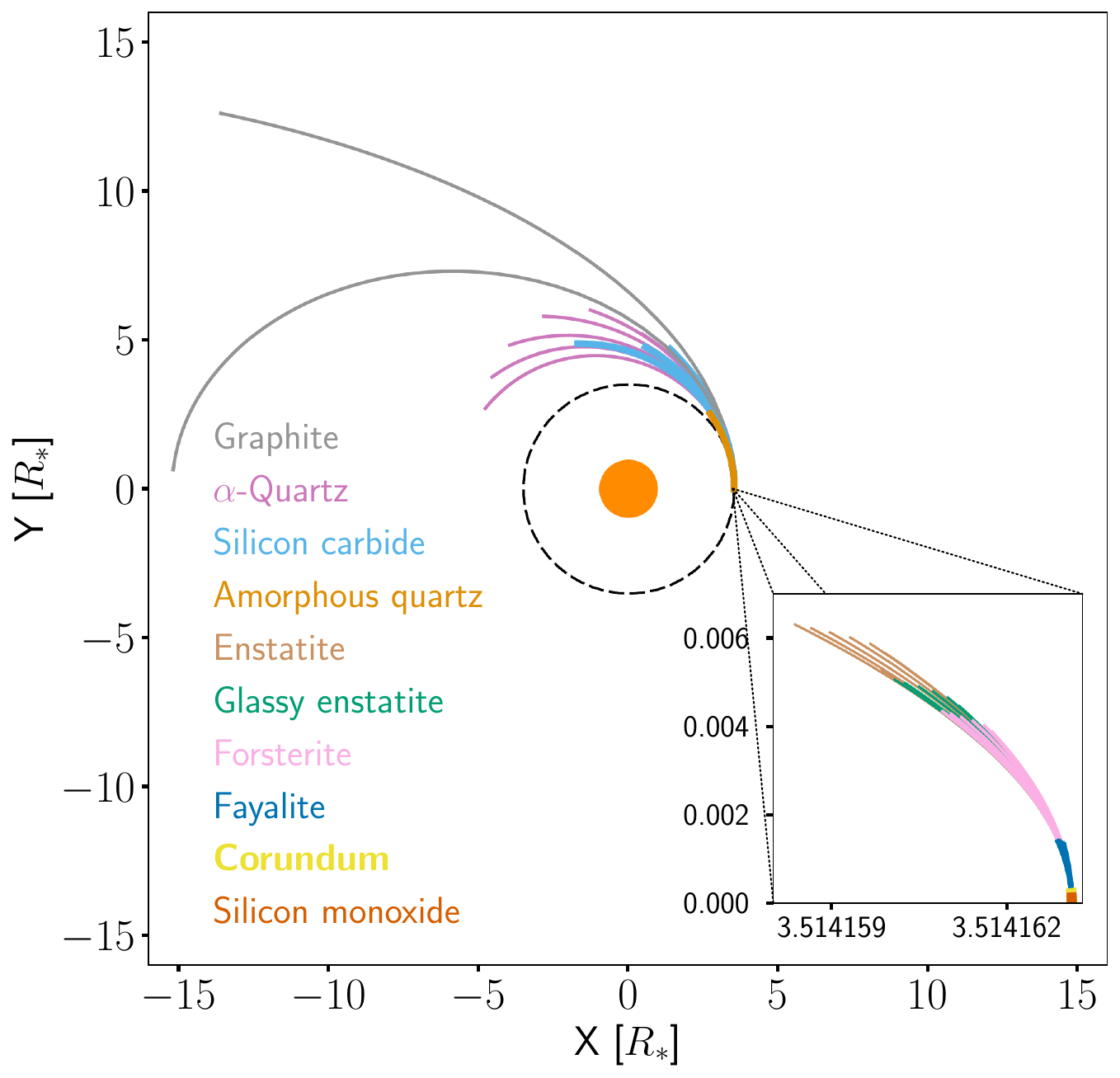}}
    \caption{Motion of dust for a selected composition subject to gravity and radiation pressure. The time interval of the trajectory is the characteristic sublimation timescale of the compound. The orange circle represents 55~Cnc~A, ajd the dashed black circle is the orbit of 55~Cnc~e scaled in units of stellar radius. The zoomed-in panel shows the trajectory of short-lived species. Different lines of the same colour represent different grain sizes of a species.}
    \label{fig:motion}
\end{figure}

\begin{figure}
    \centering
    \resizebox{\hsize}{!}{\includegraphics{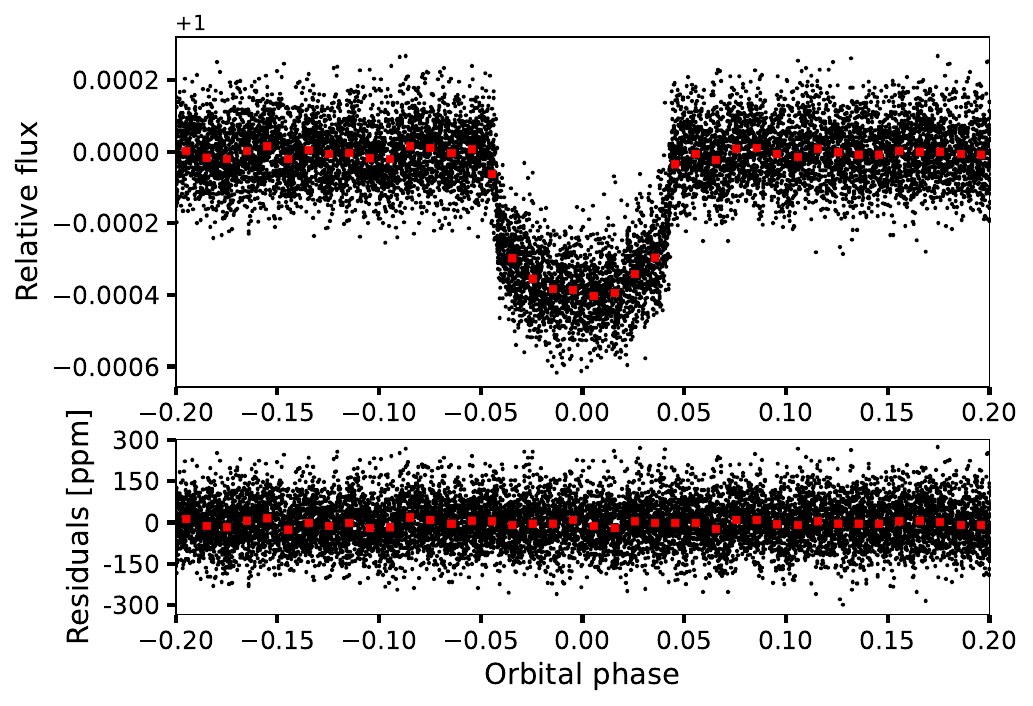}}
    \caption{Detrended flux measurements of 55 Cnc e, phase-folded at the planet orbital period. The top panel shows the relative flux, and the bottom panel presents the residuals in ppm. The plots show a portion of the phase centred around the transit. The red squares represent the binned data.}
    \label{fig:phase-fold}
\end{figure}

For the grain radii that lead to a Keplerian orbit, we computed the eccentricity and semi-major axis using Eq. (\ref{eqn:eccentricity and semimajor axis}) to obtain the trajectory of a grain during its characteristic sublimation timescale after escaping the planet. A grain of a specific species and size with periastron at the location it escaped the planetary Hill sphere radius initially moves at the same speed as the planet \citep{Rappaport_2012}. It is fair to assume this initial speed value because the ratio of the escape speed to the orbital speed is approximately $\sqrt{(M_{p}a_{p})/(M_{\star}R_{p})} \approx$ 0.01. A higher ratio of the radiation pressure to gravity leads to more eccentric and longer orbits. Figure \ref{fig:motion} shows the motion of the dust during its characteristic sublimation lifetime. Based on the timescales, most species will stay bounded to the planet or follow closely behind, forming a comet-like tail. So far, there is no evidence of a comet-like transit shape in observations of 55~Cnc~e. No asymmetry in the transit shape is observed in the CHEOPS observations, as shown in the phase-folded light curve in Fig. \ref{fig:phase-fold}.

Graphite, $\alpha$-quartz, and silicon carbide travel at least one-fourth of an entire orbit during their lifetime. Whilst the process of supplying material could be stochastic in nature, if the dust is replenished frequently enough, an inhomogeneous torus around the star could form. It would consist of regions that are more densely packed than others, and in theory, it would vary in opacity, producing measurable flux variation. While this could explain the variability in the phase-curve amplitude, it remains to be seen why this is not observed during transit. The material could be an inefficient forward-scatterer but might efficiently scatter at other angles. However, the asymmetry parameter of our considered species indicates that forward scattering mostly dominates in the CHEOPS bandpass for most of the grain sizes. If the ejecta of material occur during transit, the material would float close behind the planet and manifest in the transit shape as a less pronounced flux increase during egress and even after egress of the planet. It is possible that none of the CHEOPS visits coincided with an event like this during transit, which might explain why we did not see a comet-like transit shape \citep{Rappaport_2012} in the data. Even more, the fact that such a transit due to a comet-like tail is not observed suggests that the replenishment of material to the torus is not an uninterrupted process.

\subsection{Star-planet interactions}
\label{subsection:spi}

Magnetic star-planet interactions have been already considered in previous works as a possible origin of the chromospheric hot spots in 55 Cnc. \citet{Folsom_2020} modelled the environment of 55 Cnc based on the magnetic field constrained by Zeeman-Doppler imaging carried out in 2017, and found that 55 Cnc e is extremely likely to orbit within the sub-Alfvénic region of the stellar wind. This implies that a direct magnetic connection can be established between planet e and 55 Cnc \citep{Saur_2013}, channelling energy from the planet vicinity to the star along the magnetic topology of the stellar environment \citep{Strugarek_2016}. \citet{Morris_2021} further considered this possibility and estimated the strongest power that could be associated with such an interaction \citep{Zarka_2007} using the data from the wind model of \citet{Folsom_2020}. They found that any star-planet magnetic interaction signal could not exceed $10^{-10}$ $L_\star$. This optimistic estimate predicts a very low power and thus rules out any detectable magnetic star-planet interaction between 55 Cnc and 55 Cnc e. 

In addition, the characteristics of the 55 Cnc system also allow us to rule out such a detection. Any magnetic star-planet interaction signal will be modulated by the orbital period of the planet and the rotation rate of the star that entrains its low corona (for an example for HD 189733, see \citet{Strugarek_2022}). 55 Cnc rotates in about 40 days \citep{Henry_2000, Fisher_2008, Bourrier_2018}, while the planet orbits in about 0.74 days. If the star possesses a large-scale magnetic topology similar to the topology it had in April 2017, that is, an inclined dipole \citep{Folsom_2020}, the location of the origin of the signal would circulate around the magnetic poles of the star as the planet orbits. Because the star rotates more slowly than the orbital motion, the signal would be visible at all times (one magnetic pole would always face Earth) and should correlate with the orbital period of 55 Cnc e. In addition to this signal, radio emission from the hypothetical magnetosphere of 55 Cnc e would also be expected, again correlated with the orbital period of the planet \citep{Turner_2021}. No such signals have been detected so far. We can therefore safely reject this hypothesis for 55 Cnc e.  

Other types of star-planet interactions could nevertheless occur between the two bodies. Tidal interactions have also been proposed as a source of enhanced stellar activity \citep[e.g.][]{Cuntz_2000, Lanza_2008}. In the case of 55 Cnc, the low mass of planet e renders this scenario nevertheless implausible. The last type of star-planet interaction that could be acting in 55 Cnc would be the infall of escaping material from the planet atmosphere down to the stellar corona, producing a detectable signal \citep[e.g.][]{Matsakos_2015}. In this case, dedicated studies (beyond the scope of the present paper; e.g. \citet{DaleyYates_2019}) are needed to assess both the energetics and the relative phase of such a signal. 

\section{Conclusions}
\label{section:conclusions}

CHEOPS observations reveal a changing phase curve at least on the orbital timescale of the planet. The phase modulation varies by up to 50\,ppm. Additionally, we found a 51.9-day period of the time-dependent phase-curve amplitude of each CHEOPS visit. The origin of this is unknown. The fact that the peak of the phase curve for some visits occur during transit or close to transit rule out the planet as the source of the signal. The results motivated a deeper study on whether dust might be the cause. Our toy model allowed us to explore possible compositions of dust. The short lifetimes of some compounds such as pyroxene and olivine, whose abundance is expected to dominate at the surface of Earth-like exoplanets, mean that they are unlikely to cause a variability in the order of hours. Only a narrow range of dust sizes of graphite, silicon carbide, and quartz satisfy the required timescale. Additionally, only certain particle sizes of these species remain candidates to form a torus around the star, as suggested by past research. An argument against forming a circumstellar torus of dust is the escape velocity of approximately 24 km/s of the planet.

Previous research attributed the puzzling observations on 55~Cnc~e to be caused by an individual phenomenon. Instead of searching for a single process, a complex model including dust dynamics, magnetohydrodynamics, and atmospheric radiative transfer should provide a self-consistent model of the planet. A model should take this into account to realistically model the motion of a grain and its influence on the observations and provide conclusive answers. 

Recently, JWST observed 55~Cnc~e in the framework of two accepted programs for Cycle 1. The first set of observations aims to study the possibility of a 3:2 spin-orbit resonance as an explanation for the variable occultation depth \citep{Brandeker_2021}. The second program will characterise the atmosphere of 55~Cnc~e via spectral features of H$_{2}$O, CO, CO$_{2}$ , and SiO \citep{Hu_2021}. These observations promise exciting new insights in the IR range. Moreover, simultaneous observations of CHEOPS and JWST, accompanied by other instruments such as a spectropolarimeter, would provide a unique opportunity for resolving the nature of this fascinating exoplanet.     

\begin{acknowledgements}
We are grateful to the anonymous referee for the careful reading and thoughtful suggestions that improved this paper. We also thank the editor, Emmanuel Lellouch, for insightful comments. Both made the submission process an enjoyable experience. We also thank the language editor for the revision of the manuscript. 
CHEOPS is an ESA mission in partnership with Switzerland with important contributions to the payload and the ground segment from Austria, Belgium, France, Germany, Hungary, Italy, Portugal, Spain, Sweden, and the United Kingdom. The CHEOPS Consortium would like to gratefully acknowledge the support received by all the agencies, offices, universities, and industries involved. Their flexibility and willingness to explore new approaches were essential to the success of this mission. 
EMV aknowledges support from the Centre for Space and Habitability (CSH). This work has been carried out within the framework of the National Centre of Competence in Research PlanetS supported by the Swiss National Science Foundation under grants 51NF40\_182901 and 51NF40\_205606. EMV acknowledge the financial support of the SNSF. EMV thanks Beatriz Campos Estrada for insightful discussion on the sublimation timescales of dust in disintegrating exoplanets. 
B.-O. D. acknowledges support from the Swiss State Secretariat for Education, Research and Innovation (SERI) under contract number MB22.00046. 
ABr was supported by the SNSA. 
S.G.S. acknowledge support from FCT through FCT contract nr. CEECIND/00826/2018 and POPH/FSE (EC). 
This project has received funding from the European Research Council (ERC) under the European Union's Horizon 2020 research and innovation programme (project {\sc Spice Dune}, grant agreement No 947634). 
LBo, VNa, IPa, GPi, RRa, GSc, VSi, and TZi acknowledge support from CHEOPS ASI-INAF agreement n. 2019-29-HH.0. 
DJB acknowledges financial support from the CSH, University of Bern. 
ML acknowledges support of the Swiss National Science Foundation under grant number PCEFP2\_194576. 
ASt acknowledges support from the PLATO/CNES grant at CEA/IRFU/DAp and the french Programme National de Planétologie (PNP).
This work was supported by FCT - Fundação para a Ciência e a Tecnologia through national funds and by FEDER through COMPETE2020 - Programa Operacional Competitividade e Internacionalizacão by these grants: UID/FIS/04434/2019, UIDB/04434/2020, UIDP/04434/2020, PTDC/FIS-AST/32113/2017 \& POCI-01-0145-FEDER- 032113, PTDC/FIS-AST/28953/2017 \& POCI-01-0145-FEDER-028953, PTDC/FIS-AST/28987/2017 \& POCI-01-0145-FEDER-028987, O.D.S.D. is supported in the form of work contract (DL 57/2016/CP1364/CT0004) funded by national funds through FCT. 
YAl acknowledges the support of the Swiss National Fund under grant 200020\_172746. 
We acknowledge support from the Spanish Ministry of Science and Innovation and the European Regional Development Fund through grants ESP2016-80435-C2-1-R, ESP2016-80435-C2-2-R, PGC2018-098153-B-C33, PGC2018-098153-B-C31, ESP2017-87676-C5-1-R, MDM-2017-0737 Unidad de Excelencia Maria de Maeztu-Centro de Astrobiologí­a (INTA-CSIC), as well as the support of the Generalitat de Catalunya/CERCA programme. The MOC activities have been supported by the ESA contract No. 4000124370. 
S.C.C.B. acknowledges support from FCT through FCT contracts nr. IF/01312/2014/CP1215/CT0004. 
XB, SC, DG, MF and JL acknowledge their role as ESA-appointed CHEOPS science team members. 
ACC acknowledges support from STFC consolidated grant numbers ST/R000824/1 and ST/V000861/1, and UKSA grant number ST/R003203/1. 
This project was supported by the CNES. 
The Belgian participation to CHEOPS has been supported by the Belgian Federal Science Policy Office (BELSPO) in the framework of the PRODEX Program, and by the University of Liège through an ARC grant for Concerted Research Actions financed by the Wallonia-Brussels Federation. 
L.D. is an F.R.S.-FNRS Postdoctoral Researcher. 
This project has received funding from the European Research Council (ERC) under the European Union’s Horizon 2020 research and innovation programme (project {\sc Four Aces}. 
grant agreement No 724427). It has also been carried out in the frame of the National Centre for Competence in Research PlanetS supported by the Swiss National Science Foundation (SNSF). DE acknowledges financial support from the Swiss National Science Foundation for project 200021\_200726. 
MF and CMP gratefully acknowledge the support of the Swedish National Space Agency (DNR 65/19, 174/18). 
DG gratefully acknowledges financial support from the CRT foundation under Grant No. 2018.2323 ``Gaseous or rocky? Unveiling the nature of small worlds''. 
M.G. is an F.R.S.-FNRS Senior Research Associate. 
MNG is the ESA CHEOPS Project Scientist and Mission Representative, and as such also responsible for the Guest Observers (GO) Programme. MNG does not relay proprietary information between the GO and Guaranteed Time Observation (GTO) Programmes, and does not decide on the definition and target selection of the GTO Programme. 
SH gratefully acknowledges CNES funding through the grant 837319. 
KGI is the ESA CHEOPS Project Scientist and is responsible for the ESA CHEOPS Guest Observers Programme. She does not participate in, or contribute to, the definition of the Guaranteed Time Programme of the CHEOPS mission through which observations described in this paper have been taken, nor to any aspect of target selection for the programme. 
This work was granted access to the HPC resources of MesoPSL financed by the Region Ile de France and the project Equip@Meso (reference ANR-10-EQPX-29-01) of the programme Investissements d'Avenir supervised by the Agence Nationale pour la Recherche. 
PM acknowledges support from STFC research grant number ST/M001040/1. 
This work was also partially supported by a grant from the Simons Foundation (PI Queloz, grant number 327127). 
IRI acknowledges support from the Spanish Ministry of Science and Innovation and the European Regional Development Fund through grant PGC2018-098153-B- C33, as well as the support of the Generalitat de Catalunya/CERCA programme. 
GyMSz acknowledges the support of the Hungarian National Research, Development and Innovation Office (NKFIH) grant K-125015, a a PRODEX Experiment Agreement No. 4000137122, the Lend\"ulet LP2018-7/2021 grant of the Hungarian Academy of Science and the support of the city of Szombathely. 
V.V.G. is an F.R.S-FNRS Research Associate. 
NAW acknowledges UKSA grant ST/R004838/1. 
ACC and TW acknowledge support from STFC consolidated grant numbers ST/R000824/1 and ST/V000861/1, and UKSA grant number ST/R003203/1. 
NCS acknowledges funding by the European Union (ERC, FIERCE, 101052347). Views and opinions expressed are however those of the author(s) only and do not necessarily reflect those of the European Union or the European Research Council. Neither the European Union nor the granting authority can be held responsible for them. 
This research made use of \texttt{exoplanet} \citep{Foreman-Mackey_2021} and its
dependencies \citep{exoplanet:agol20, exoplanet:arviz, exoplanet:astropy13,
exoplanet:astropy18, exoplanet:kipping13, exoplanet:luger18, Salvatier_2016,
exoplanet:theano}. We acknowledge the use of further software: \texttt{batman} \citep{Kreidberg_2015}, \texttt{NumPy} \citep{Harris_2020}, \texttt{matplotlib} \citep{Hunter_2007}, \texttt{corner} \citep{corner_2016}, \texttt{astroquery} \citep{Ginsburg_2019}, \texttt{seaborn} \citep{Waskom_2021} and \texttt{scipy} \citep{scipy_2020}. 

\end{acknowledgements}

\bibliographystyle{aa}
\setcitestyle{authoryear,open={(},close={)}}
\bibliography{reference}

\begin{appendix}
\label{section_appendix}

\section{Detrending basis vectors}
\label{appendix:basis vectors}

Table \ref{tab:basis vectors} provides information of the basis vectors we used to correct for systematics from the flux measurements in each CHEOPS visit. The selection is based on the combination of basis vectors that minimise the BIC. 

\begin{table*}[!hbt]
\centering\setstretch{1.5}
\caption{Basis vectors used to detrend each visit.}
\begin{tabular}{c|c|c|c|c|c|c|c|c|c|c|c|c|c|c}
\hline
\hline
Visit & $cos(\psi)$ & $sin(\psi)$ & $cos(2\psi)$ & $sin(2\psi)$ & $cos(3\psi)$ & $sin(3\psi)$ & $cos(4\psi)$ & $sin(4\psi)$ & $t$ & $t^{2}$ & BG & BG$^{2}$ & tF\_2 & tF\_2$^{2}$ \\
\hline 
1 & \checkmark & \checkmark & \checkmark & \checkmark & & & & & \checkmark & & \checkmark &  & \checkmark & \\
\hline 
2 & \checkmark & \checkmark & \checkmark & \checkmark & & & & & \checkmark & \checkmark & & & \checkmark & \\
\hline 
3 & \checkmark & \checkmark & \checkmark & \checkmark & & & & & \checkmark & \checkmark & & & \checkmark & \\
\hline 
4 & \checkmark & \checkmark & \checkmark & \checkmark & & & & & \checkmark & \checkmark & & & \checkmark & \\
\hline 
5 & \checkmark & \checkmark & \checkmark & \checkmark & & & & & \checkmark & \checkmark & & & \checkmark & \\
\hline 
6 & \checkmark & \checkmark & \checkmark & \checkmark & & & & & \checkmark & & & & \checkmark & \\
\hline 
7 & \checkmark & \checkmark & \checkmark & \checkmark & & & & & \checkmark & & & & \checkmark & \\
\hline 
8 & \checkmark & \checkmark & \checkmark & \checkmark & & & & & \checkmark & \checkmark & \checkmark & & \checkmark & \\
\hline 
9 & \checkmark & \checkmark & \checkmark & \checkmark & & & & & \checkmark & \checkmark & \checkmark & & \checkmark & \\
\hline 
10 & \checkmark & \checkmark & \checkmark & \checkmark & & & & & \checkmark & \checkmark & \checkmark & & \checkmark & \\
\hline 
11 & \checkmark & \checkmark & \checkmark & \checkmark & & & & & \checkmark & \checkmark & & & \checkmark & \\
\hline 
12 & \checkmark & \checkmark & \checkmark & \checkmark & & & & & \checkmark & & \checkmark & & \checkmark & \\
\hline 
13 & \checkmark & \checkmark & \checkmark & \checkmark & & & & & \checkmark & \checkmark & \checkmark & & \checkmark & \\
\hline 
14 & \checkmark & \checkmark & \checkmark & \checkmark & & & & & \checkmark & & \checkmark & & \checkmark & \\
\hline 
15 & \checkmark & \checkmark & \checkmark & \checkmark & & & & & \checkmark & \checkmark & \checkmark & & \checkmark & \\
\hline 
16 & \checkmark & \checkmark & \checkmark & \checkmark & & & & & \checkmark & \checkmark & & & \checkmark & \\
\hline 
17 & \checkmark & \checkmark & \checkmark & \checkmark & & & & & \checkmark & & \checkmark & & \checkmark & \\
\hline 
18 & \checkmark & \checkmark & \checkmark & \checkmark & & & & & \checkmark & \checkmark & \checkmark & & \checkmark & \\
\hline 
19 & \checkmark & \checkmark & \checkmark & \checkmark & & & & & \checkmark & \checkmark & \checkmark & & \checkmark & \\
\hline 
20 & \checkmark & \checkmark & \checkmark & \checkmark & & & & & \checkmark & & \checkmark & & \checkmark & \\
\hline 
21 & \checkmark & \checkmark & \checkmark & \checkmark & & & & & \checkmark & & \checkmark & & \checkmark & \\
\hline 
22 & \checkmark & \checkmark & \checkmark & \checkmark & & & & & \checkmark & \checkmark & \checkmark & & \checkmark & \\
\hline 
23 & \checkmark & \checkmark & \checkmark & \checkmark & & & & & \checkmark & \checkmark & \checkmark & & \checkmark & \checkmark \\
\hline 
24 & \checkmark & \checkmark & \checkmark & \checkmark & & & & & \checkmark & \checkmark & \checkmark & & \checkmark & \checkmark \\
\hline 
25 & \checkmark & \checkmark & \checkmark & \checkmark & & & & & \checkmark & \checkmark & & & \checkmark & \\
\hline 
26 & \checkmark & \checkmark & \checkmark & \checkmark & & & & & \checkmark & \checkmark & & & \checkmark & \\
\hline 
27 & \checkmark & \checkmark & \checkmark & \checkmark & \checkmark & \checkmark & \checkmark & \checkmark & \checkmark & \checkmark & \checkmark & \checkmark & \checkmark & \\
\hline 
28 & \checkmark & \checkmark & \checkmark & \checkmark & & & & & \checkmark & \checkmark & \checkmark & & \checkmark & \checkmark \\
\hline 
29 & \checkmark & \checkmark & \checkmark & \checkmark & & & & & \checkmark & & \checkmark & \checkmark & \checkmark & \\
\hline
\end{tabular}
\tablefoot{
 A blank space indicates that the corresponding basis vector was not used. The reported value is the posterior distribution mean and $1\sigma$ uncertainty. $\psi$ stands for the roll angle, $t$ stands for time, BG stands for background level, and tF\_2 is the thermFront\_2 thermistor readout.
}
\label{tab:basis vectors}
\end{table*}
\FloatBarrier

\section{Residuals}
\label{appendix:residuals}

\begin{figure*}[!h]
\centering
\subfigure{\includegraphics[width=0.85\textwidth]{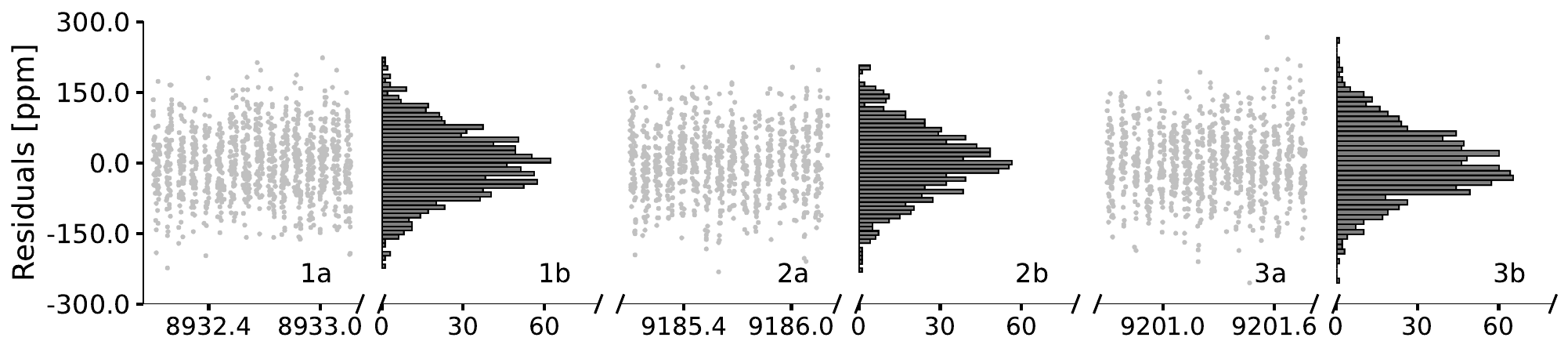}}

\vspace{-6mm}
\subfigure{\includegraphics[width=0.85\textwidth]{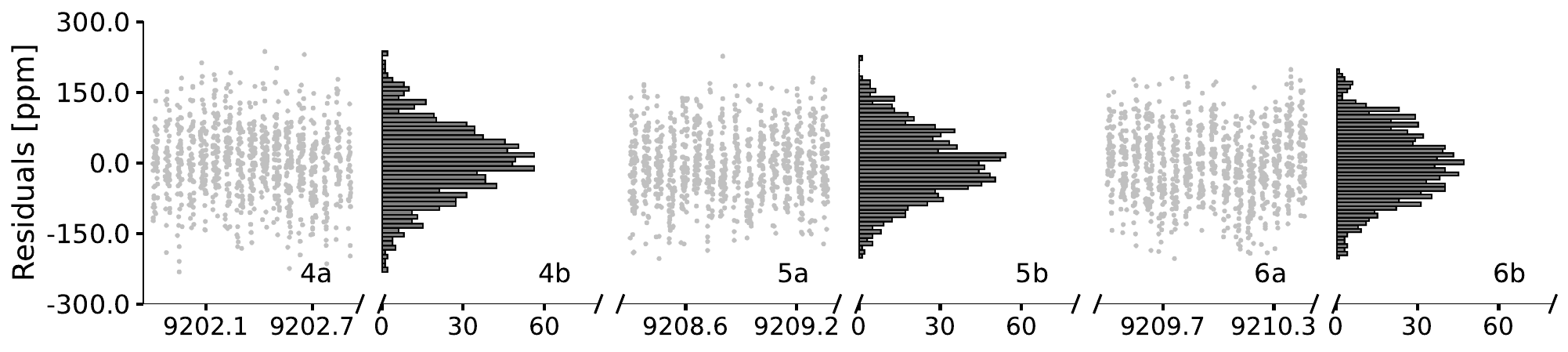}}

\vspace{-6mm}
\subfigure{\includegraphics[width=0.85\textwidth]{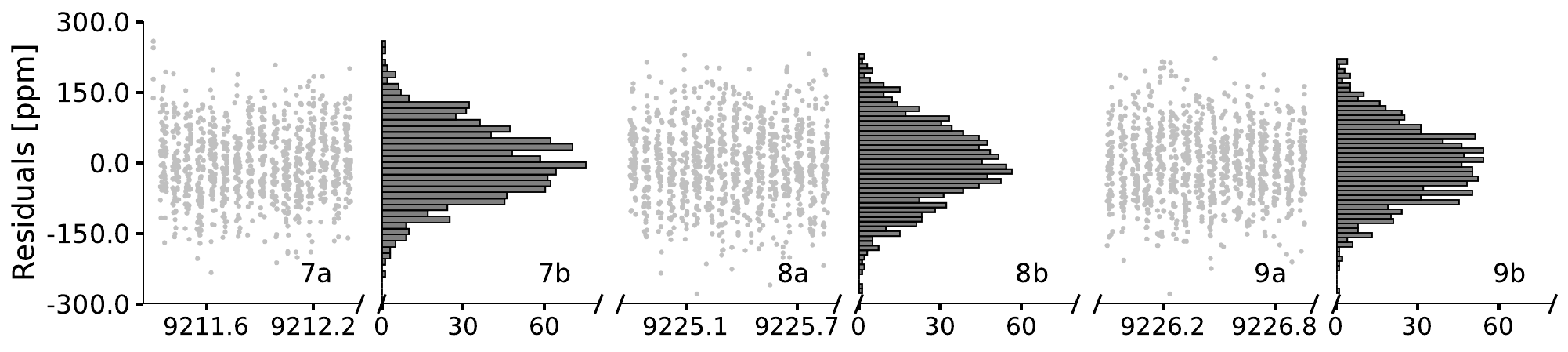}}

\vspace{-6mm}
\subfigure{\includegraphics[width=0.85\textwidth]{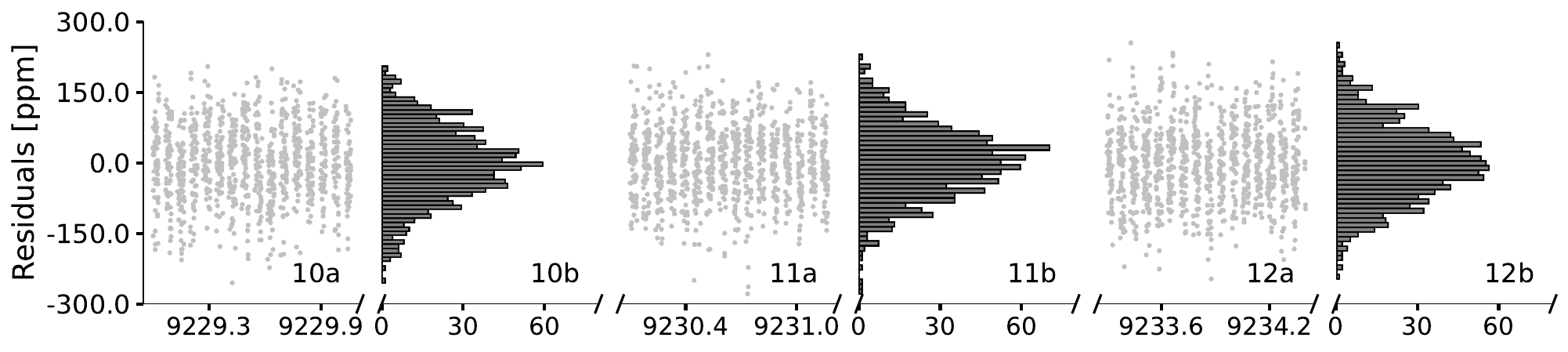}}

\vspace{-6mm}
\subfigure{\includegraphics[width=0.85\textwidth]{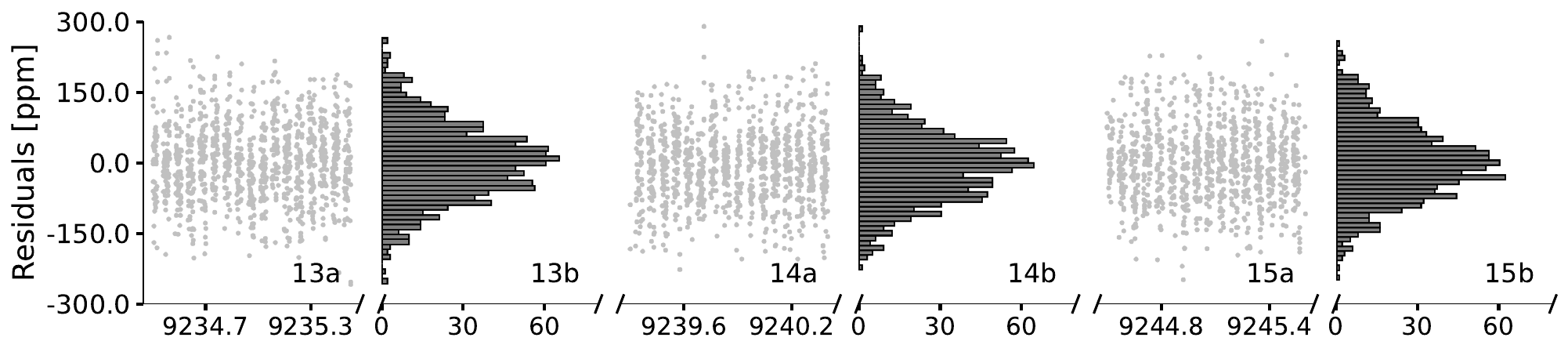}}

\caption{Residual flux in ppm assuming a sinusoidal shape of the phase curve for CHEOPS visits 1 to 15. The number in each panel corresponds to the CHEOPS visit, where the subscript a) refers to the residuals vs. BJD time-2450000 in the x-axis. The subscript b) refers to the density distribution, with density in the x-axis and residuals in the y-axis. All panels share the same y-axis scale.}
\label{fig:gallery residuals 1}
\end{figure*}

\begin{figure*}[!h]
\centering
\subfigure{\includegraphics[width=0.85\textwidth]{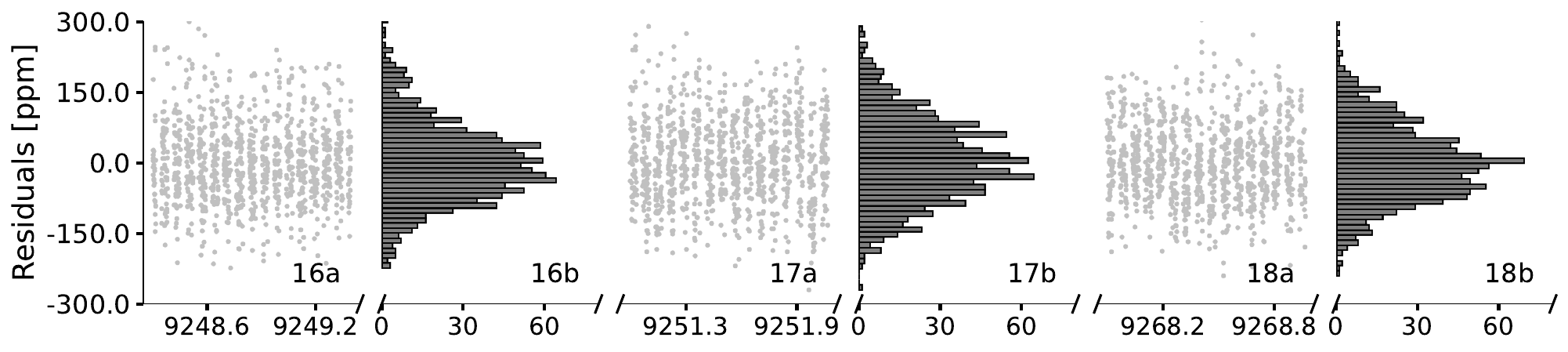}}

\vspace{-6mm}
\subfigure{\includegraphics[width=0.85\textwidth]{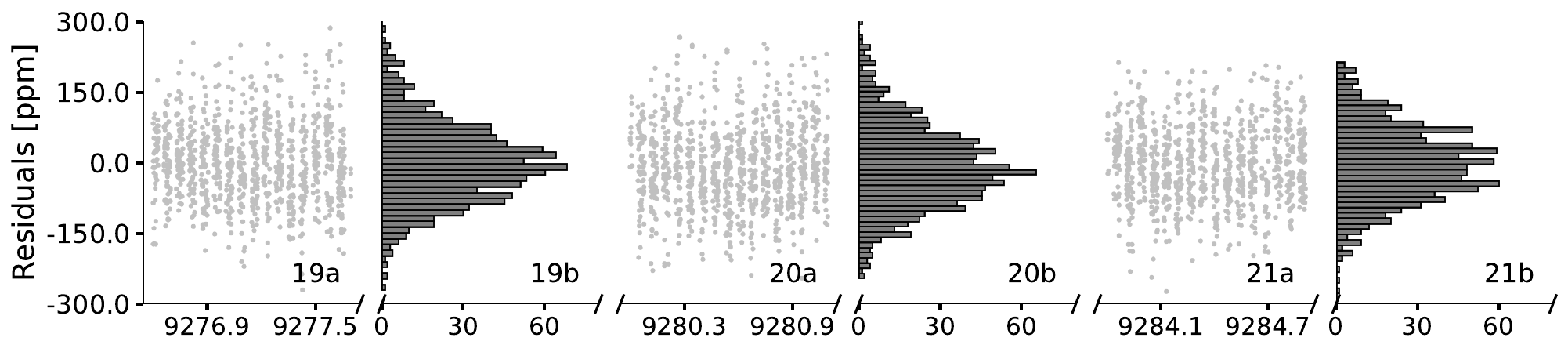}}

\vspace{-6mm}
\subfigure{\includegraphics[width=0.85\textwidth]{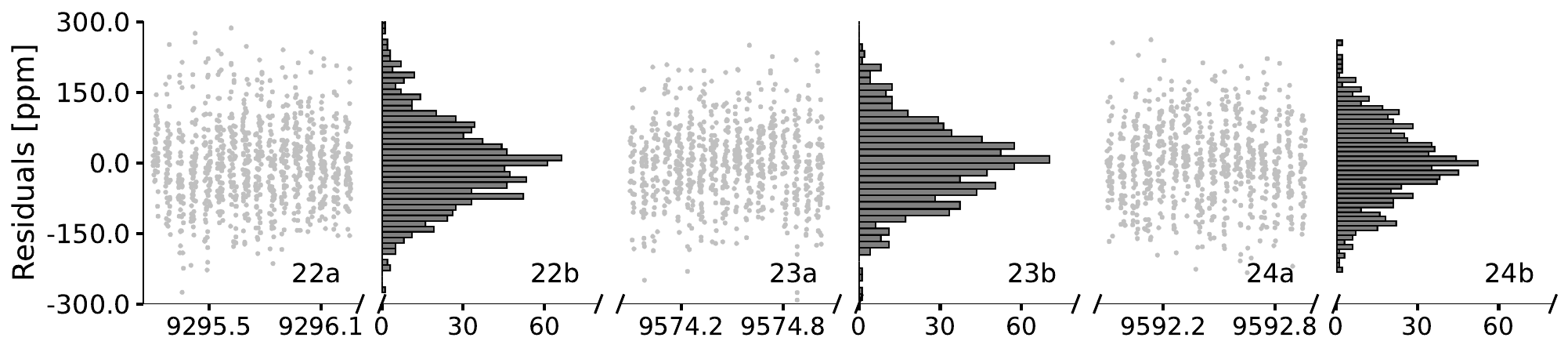}}

\vspace{-6mm}
\subfigure{\includegraphics[width=0.85\textwidth]{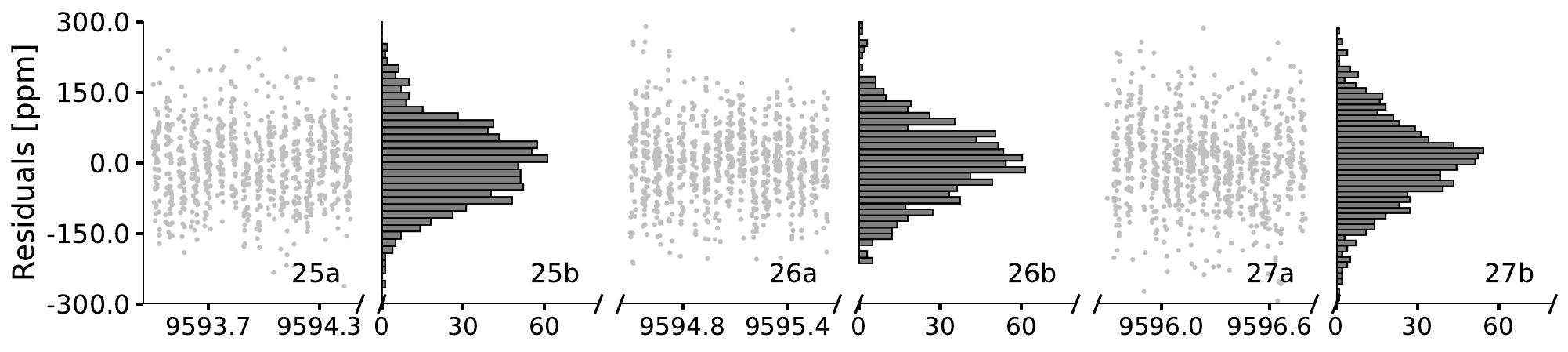}}

\vspace{-6mm}
\subfigure{\includegraphics[width=0.85\textwidth]{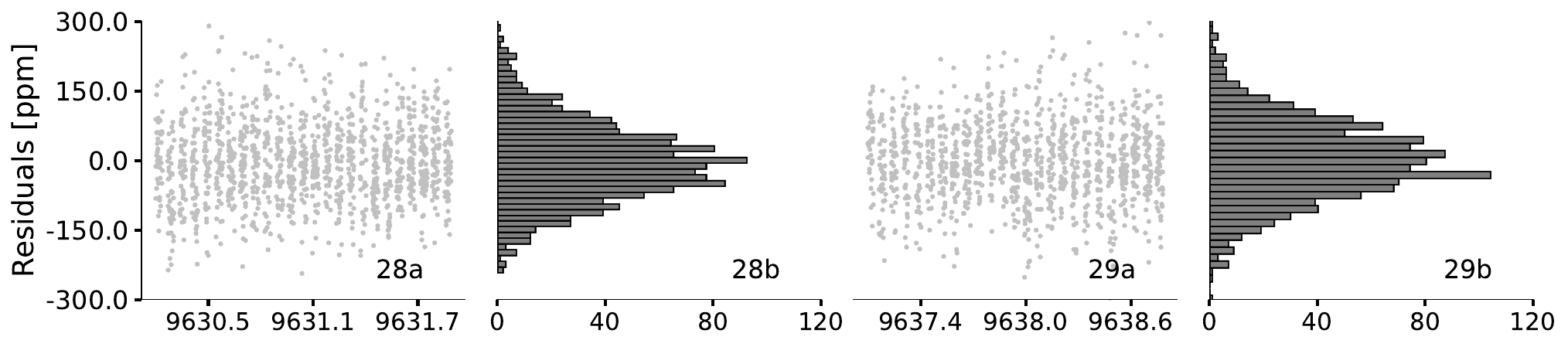}}

\caption{Same as Fig. \ref{fig:gallery residuals 1} for CHEOPS visits 16 to 29.}
\label{fig:gallery residuals 2}
\end{figure*}

The residuals correspond to the flux after systematics, transit, occultation model, and the sinusoidal signal of the phase curve are removed. The gallery in Figs. \ref{fig:gallery residuals 1} and \ref{fig:gallery residuals 2} contains the residuals of all CHEOPS visits. Complementary to this, the density distribution of the residuals is shown next to the time series. 

\FloatBarrier

\FloatBarrier

\section{Phase-curve parameters}
\label{appendix:phase curve parameters}

\begin{figure}[h!]
    \centering
    \resizebox{0.7\hsize}{!}{\includegraphics{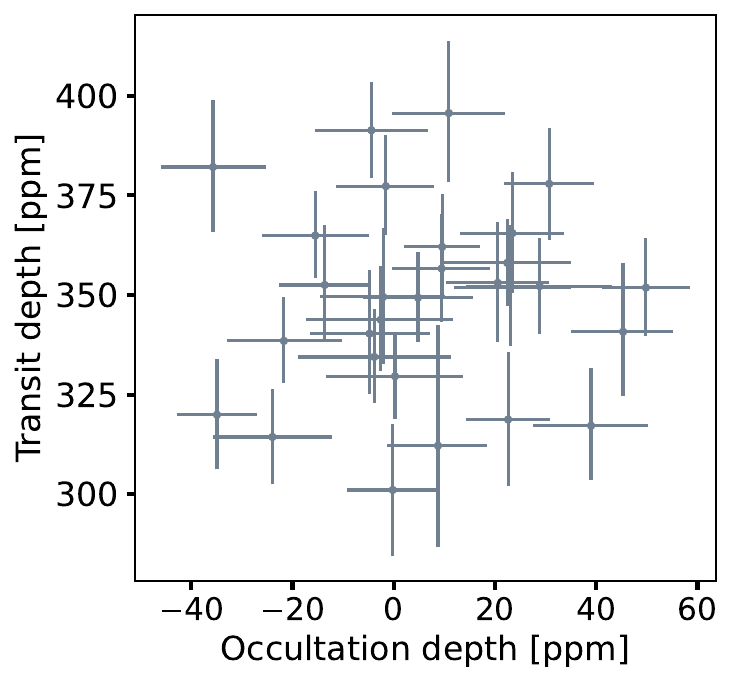}}
    \caption{Transit depth vs. occultation depth.}
    \label{fig:transit-occultation}
\end{figure}

\begin{figure}[h!]
    \centering
    \resizebox{0.7\hsize}{!}{\includegraphics{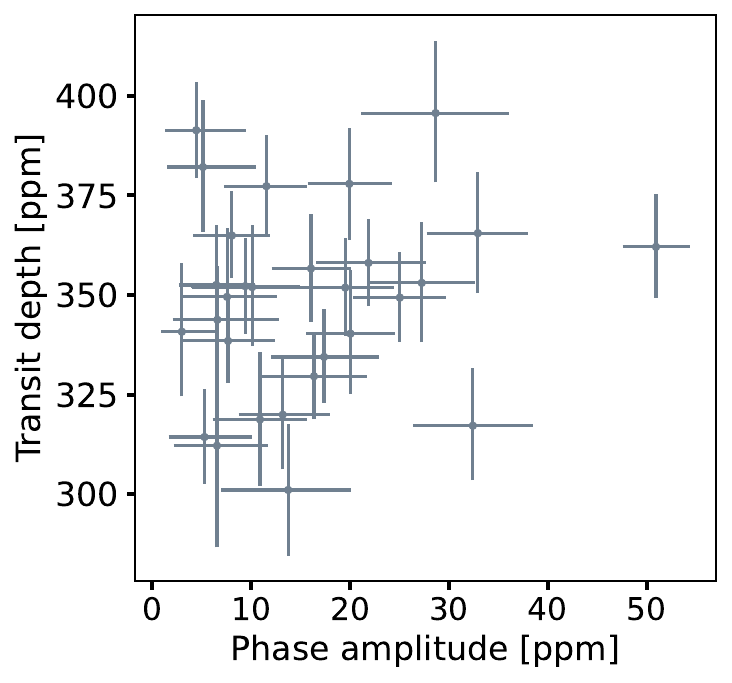}}
    \caption{Transit depth vs. phase-curve amplitude.}
    \label{fig:transit-amplitude}
\end{figure}

\begin{figure}[h!]
    \centering
    \resizebox{0.7\hsize}{!}{\includegraphics{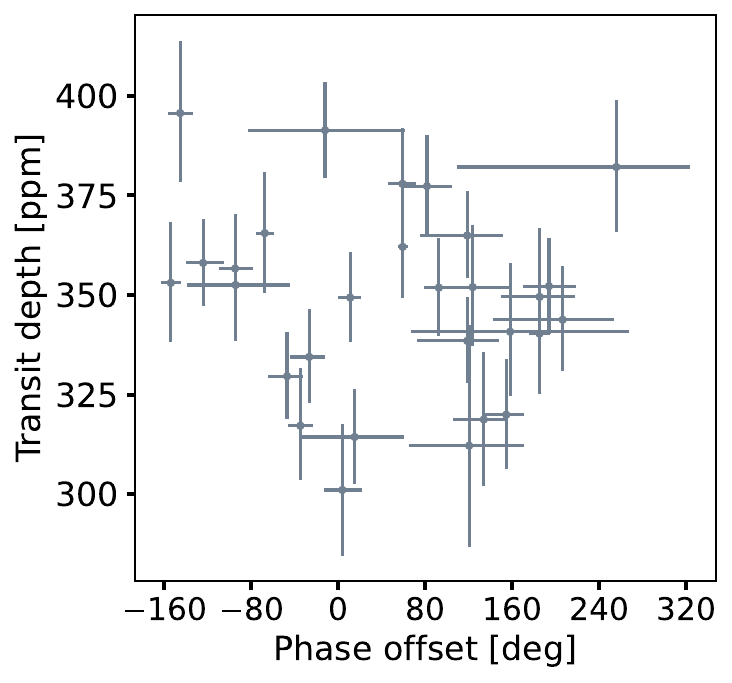}}
    \caption{Transit depth vs. phase offset.}
    \label{fig:transit-offset}
\end{figure}

\begin{figure}[h!]
    \centering
    \resizebox{0.7\hsize}{!}{\includegraphics{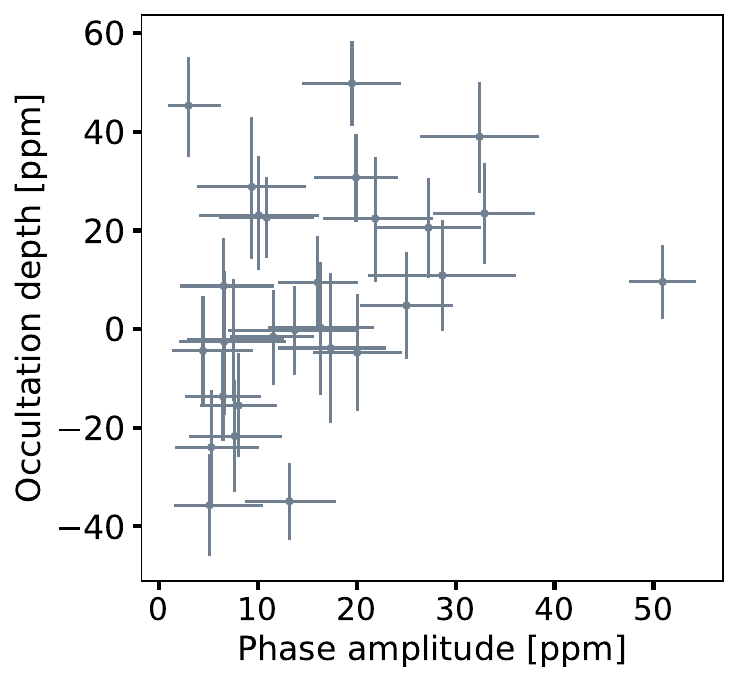}}
    \caption{Occultation depth vs. phase-curve amplitude.}
    \label{fig:occultation-amplitude}
\end{figure}

\begin{figure}[h!]
    \centering
    \resizebox{0.7\hsize}{!}{\includegraphics{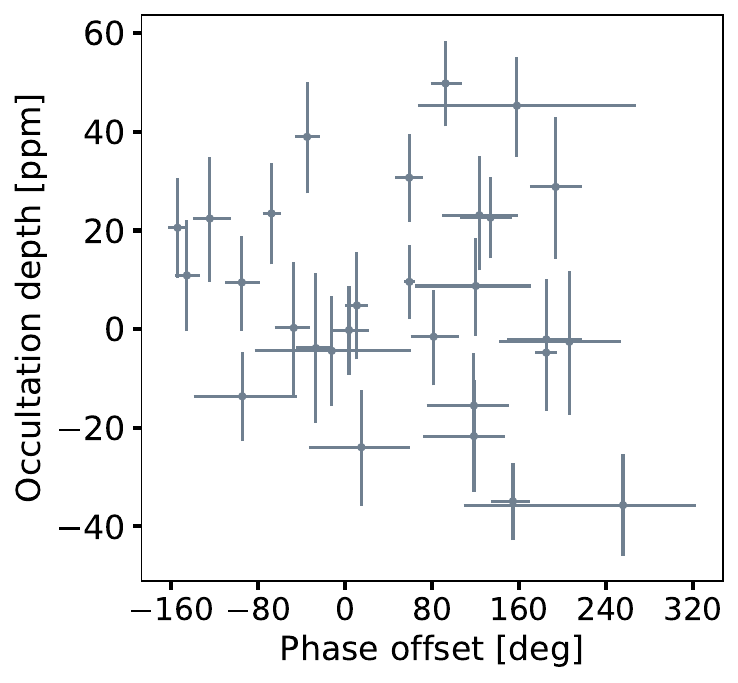}}
    \caption{Occultation depth vs. phase offset.}
    \label{fig:occultation-offset}
\end{figure}

\begin{figure}[h!]
    \centering
    \resizebox{0.7\hsize}{!}{\includegraphics{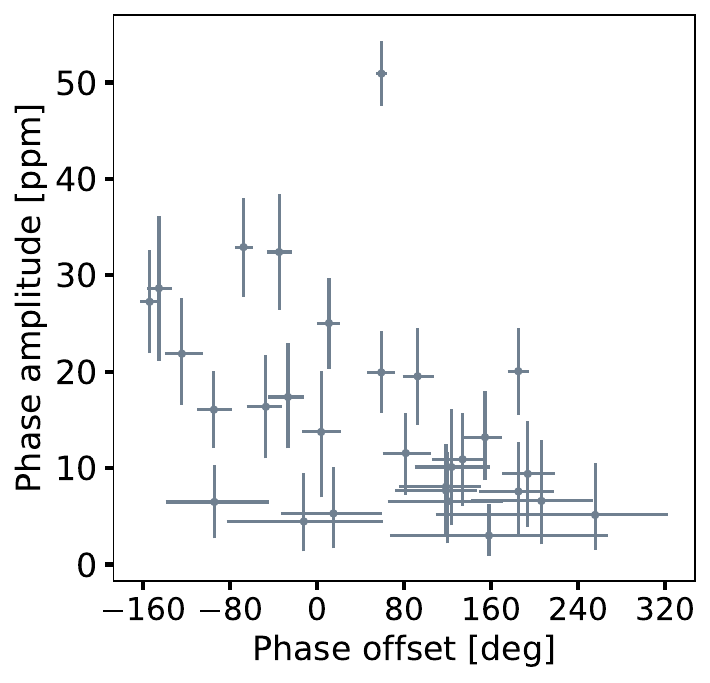}}
    \caption{Phase-curve amplitude vs. phase offset.}
    \label{fig:amplitude-offset}
\end{figure}

Here we present pairwise relations between relevant phase-curve parameters: transit and occultation depth, and phase-curve amplitude and offset. The transit depth is compared with the occultation depth (Fig. \ref{fig:transit-occultation}), phase-curve amplitude (Fig. \ref{fig:transit-amplitude}), and phase offset (Fig. \ref{fig:transit-offset}). Then the occultation depth is compared with amplitude (Fig. \ref{fig:occultation-amplitude}) and offset (Fig. \ref{fig:occultation-offset}). The remaining combination relates the phase-curve amplitude with the offset (Fig. \ref{fig:amplitude-offset}).

\FloatBarrier

\FloatBarrier

\FloatBarrier

\FloatBarrier

\FloatBarrier

\FloatBarrier

\section{Selected corner plots}
\label{appendix:corner plots}

\begin{figure*}
    \centering
    \resizebox{\hsize}{!}{\includegraphics{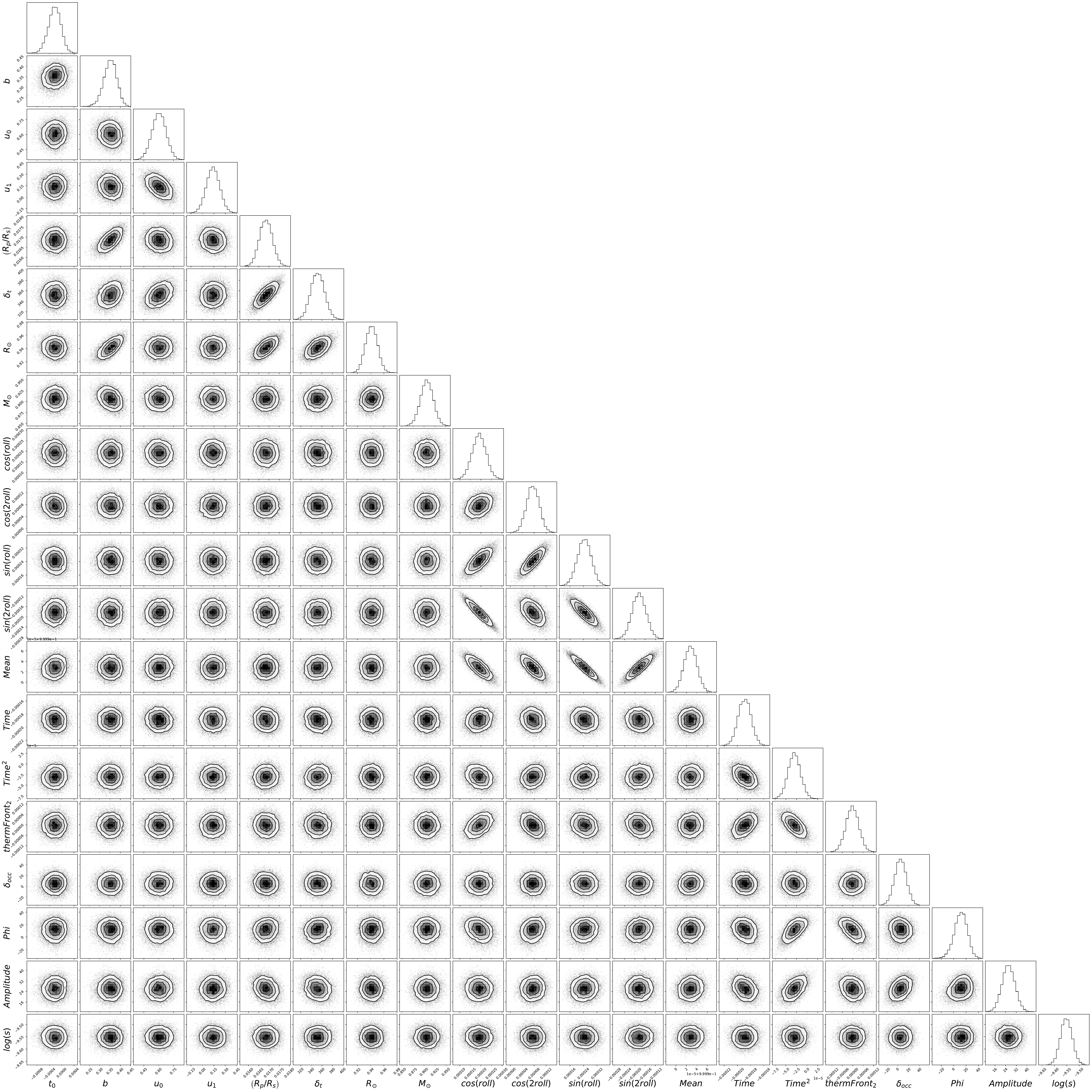}}
    \caption{Posterior distributions and joint correlations plot corresponding to CHEOPS visit 2.}
    \label{fig:corner 2}
\end{figure*}

\begin{figure*}
    \centering
    \resizebox{\hsize}{!}{\includegraphics{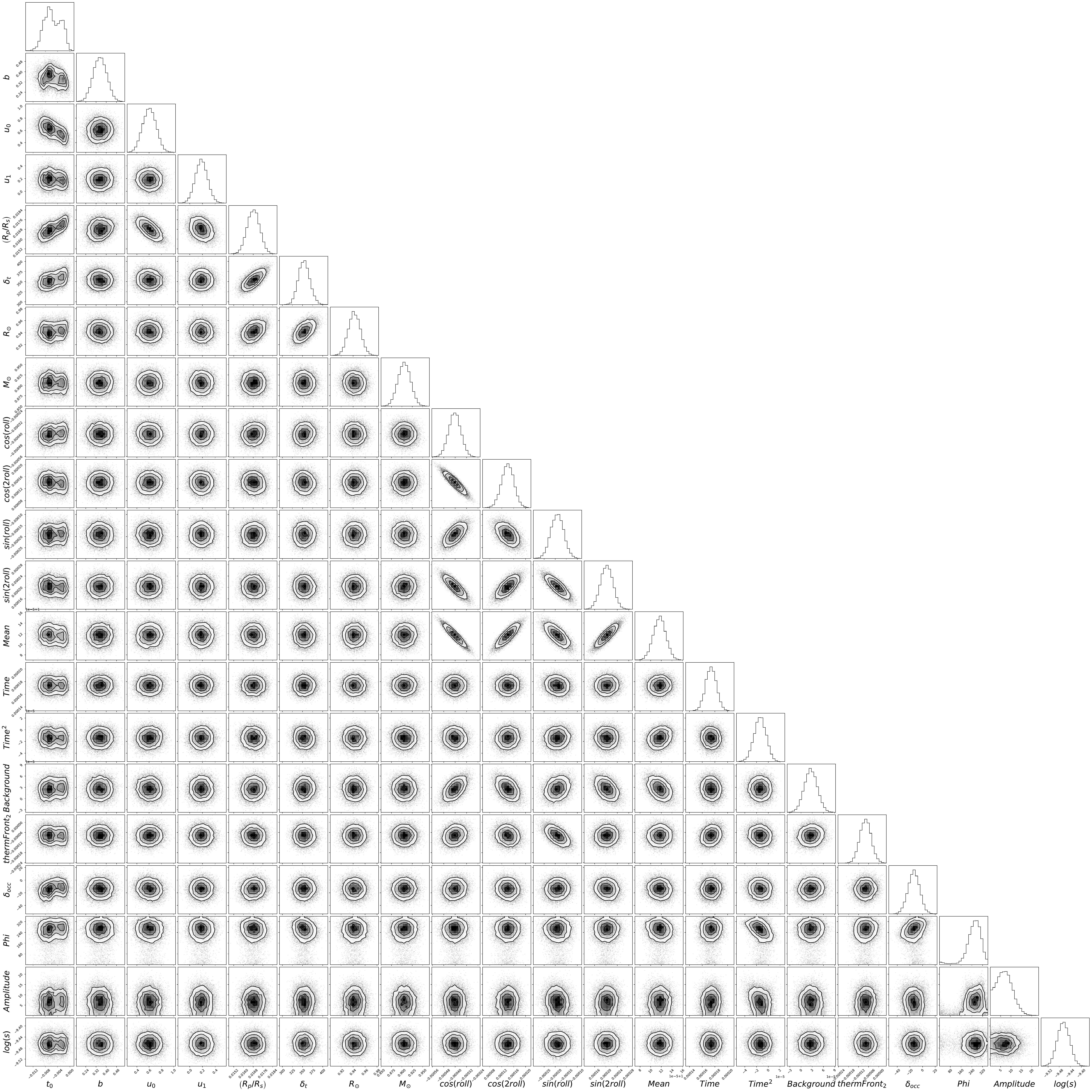}}
    \caption{Same as Fig. \ref{fig:corner 2}, but corresponding to CHEOPS visit 9.}
    \label{fig:corner 9}
\end{figure*}

Figure \ref{fig:corner 2} and \ref{fig:corner 9} show the full joint posterior correlation plots of visits 2 and 9, respectively. $b$ is the impact parameter, $t_{0}$ is the mid-transit time relative to the reference time, $R_{p}/R_{\star}$ is the ratio of the planetary radius to the stellar radius, $(R_{p}/R_{\star})^{2}$ is the transit depth, $\delta_{t}$ is the analytical expression of the transit depth \citep{Heller_2019}, $R_{\odot}$ and $M_{\odot}$ are the stellar radius and mass, $roll$ refers to the roll angle, $\delta_{occ}$ is the occultation depth, $Phi$ is the phase offset in degrees, and $\log(s)$ is the natural logarithm of the flux uncertainty for each measurement.
Visit 9 clarifies the nature of the high uncertainty in the phase offset reported in Table \ref{tab:phase curve summary}. Since ingress or egress were not observed by CHEOPS, the MCMC infers a bimodal distribution on the mid-transit time. 

\FloatBarrier

\FloatBarrier

\section{Model comparison}
\label{appendix:model comparison}

For each visit, we performed a comparison between different models for the phase curve: sinusoid function, Lambertian sphere, piecewise-Lambertian, constant baseline flux, and constant baseline flux without occultation (set to zero). Table \ref{tab:model comparison} shows the top-ranked model and the difference in the LOO between the first and second ranked model. The last column is the statistical weight of the top-ranked model. 

\begin{table}[h]
\centering\setstretch{1.5}
\caption{Statistical weight of the preferred model for each visit based on the LOO.}
\begin{tabular}{cccc}
\hline
\hline
Visit & Preferred model & $\Delta$LOO & Weight \\
\hline 
1 & Sinusoid & 51.16 & 0.83 \\
2 & Sinusoid & 9.45 & 0.82 \\
3 & Sinusoid & 11.38 & 0.84 \\
4 & Sinusoid & 1.02 & 0.63 \\
5 & Sinusoid & 0.91 & 0.50 \\
6 & Sinusoid & 7.42 & 0.61 \\
7 & Sinusoid & 6.57 & 0.86 \\
8 & Flat phase curve & 1.38 & 0.99 \\
9 & Flat phase curve & 1.01 & 0.99 \\
10 & Sinusoid & 14.85 & 0.99 \\
11 & Flat phase curve & 0.43 & 0.56 \\
12 & Sinusoid & 1.64 & 0.72 \\
13 & Sinusoid & 1.54 & 0.41 \\
14 & Sinusoid & 6.19 & 0.77 \\
15 & Sinusoid & 5.23 & 0.99 \\
16 & Sinusoid & 10.25 & 0.83 \\
17 & Sinusoid & 20.68 & 0.91 \\
18 & Sinusoid & 3.66 & 0.84 \\
19 & Flat phase curve without eclipse & 0.46 & 0.56 \\
20 & Sinusoid & 4.78 & 0.80 \\
21 & Flat phase curve & 0.61 & 0.78 \\
22 & Piecewise-Lambertian & 7.64 & 0.96 \\
23 & Flat phase curve & 0.69 & 0.99 \\
24 & Flat phase curve without eclipse & 1.04 & 0.99 \\
25 & Flat phase curve & 0.25 & 0.57 \\
26 & Sinusoid & 2.10 & 0.66 \\
27 & Sinusoid & 4.74 & 0.83 \\
28 & Flat phase curve without eclipse & 12.12 & 0.77 \\
29 & Flat phase curve without eclipse & 0.83 & 0.69 \\
\hline
\end{tabular}
\tablefoot{
The reported difference in the LOO is relative to the model ranked second.
}
\label{tab:model comparison}
\end{table}
\FloatBarrier

\end{appendix}

\end{document}